\documentclass[journal=ancac3,manuscript=article,layout=twocolumn]{achemso}

\setkeys{acs}{articletitle = true} 
\usepackage{natbib}
\usepackage{chemformula} 
\usepackage[T1]{fontenc} 
\usepackage{float} %
\usepackage{seqsplit} 
\usepackage{graphicx}
\usepackage{booktabs} 
\usepackage{multirow}
\usepackage{soul}
\usepackage{xcolor}
\usepackage{color, colortbl}
\usepackage{threeparttable}
\usepackage{ulem}
\usepackage[super]{nth}
\usepackage{amsmath}
\usepackage[font=small,labelfont=bf]{caption}
\usepackage[switch]{lineno}
\usepackage{breqn}
\usepackage{makecell} 
\usepackage{amssymb}

\author{Binay P. Nayak}
\affiliation{Ames National Laboratory, and Department of Chemical and Biological Engineering, Iowa State University, Ames, Iowa 50011, United States }

\author{James Ethan Batey}
\affiliation{Division of Materials Sciences and Engineering, Ames National Laboratory, U.S. DOE, Ames, Iowa 50011, United States}
\altaffiliation{Department of Chemistry and Biochemistry, University of Arkansas, Fayetteville, Arkansas 72701, United States}

\author{Hyeong Jin Kim}
\affiliation{Ames National Laboratory, and Department of Chemical and Biological Engineering, Iowa State University, Ames, Iowa 50011, United States }

\author{Wenjie Wang}
\affiliation{Division of Materials Sciences and Engineering, Ames National Laboratory, U.S. DOE, Ames, Iowa 50011, United States}

\author{Wei Bu}
\affiliation{NSF’s ChemMatCARS, Pritzker School of Molecular Engineering, University of Chicago, Chicago, Illinois 60637, United States}

\author{Honghu Zhang}
\affiliation{Center for Functional Nanomaterials, Brookhaven National Laboratory, Upton, New York 11973, United States}
\altaffiliation{National Synchrotron Light Source II, Brookhaven National Laboratory, Upton,
New York 11973, United States}

\author{Surya K. Mallapragada}
\email{suryakm@iastate.edu}
\affiliation{Ames National Laboratory, and Department of Chemical and Biological Engineering, Iowa State University, Ames, Iowa 50011, United States }

\author{David Vaknin}
\email{vaknin@ameslab.gov}
\affiliation{Ames National Laboratory, and Department of Physics and Astronomy, Iowa State University, Ames, Iowa 50011, United States}

\title
{Effect of Grafting Density on the Two-dimensional Assembly of Nanoparticles}

\begin{document}

\begin{abstract}

Employing grazing-incidence small-angle X-ray scattering (GISAXS) and X-ray reflectivity (XRR), we demonstrate that films composed of polyethylene glycol (PEG)-grafted silver nanoparticles (AgNP) and gold nanoparticles (AuNP), as well as their binary mixtures, form highly stable hexagonal structures at the vapor-liquid interface. These nanoparticles exhibit remarkable stability under varying environmental conditions, including changes in pH, mixing concentration, and PEG chain length. Short-chain PEG grafting produces dense, well-ordered films, while longer chains produce more complex, less dense quasi-bilayer structures. AuNPs exhibit higher grafting densities than AgNPs, leading to more ordered in-plane arrangements. In binary mixtures, AuNPs dominate the population at the surface, while AgNPs integrate into the system, expanding the lattice without forming a distinct binary superstructure. These results offer valuable insights into the structural behavior of PEG-grafted nanoparticles and provide a foundation for optimizing binary nanoparticle assemblies for advanced nanotechnology applications.
\end{abstract}

\section{Graphical Abstarct}
\begin{figure}
    \centering
    \includegraphics[width=1\linewidth]{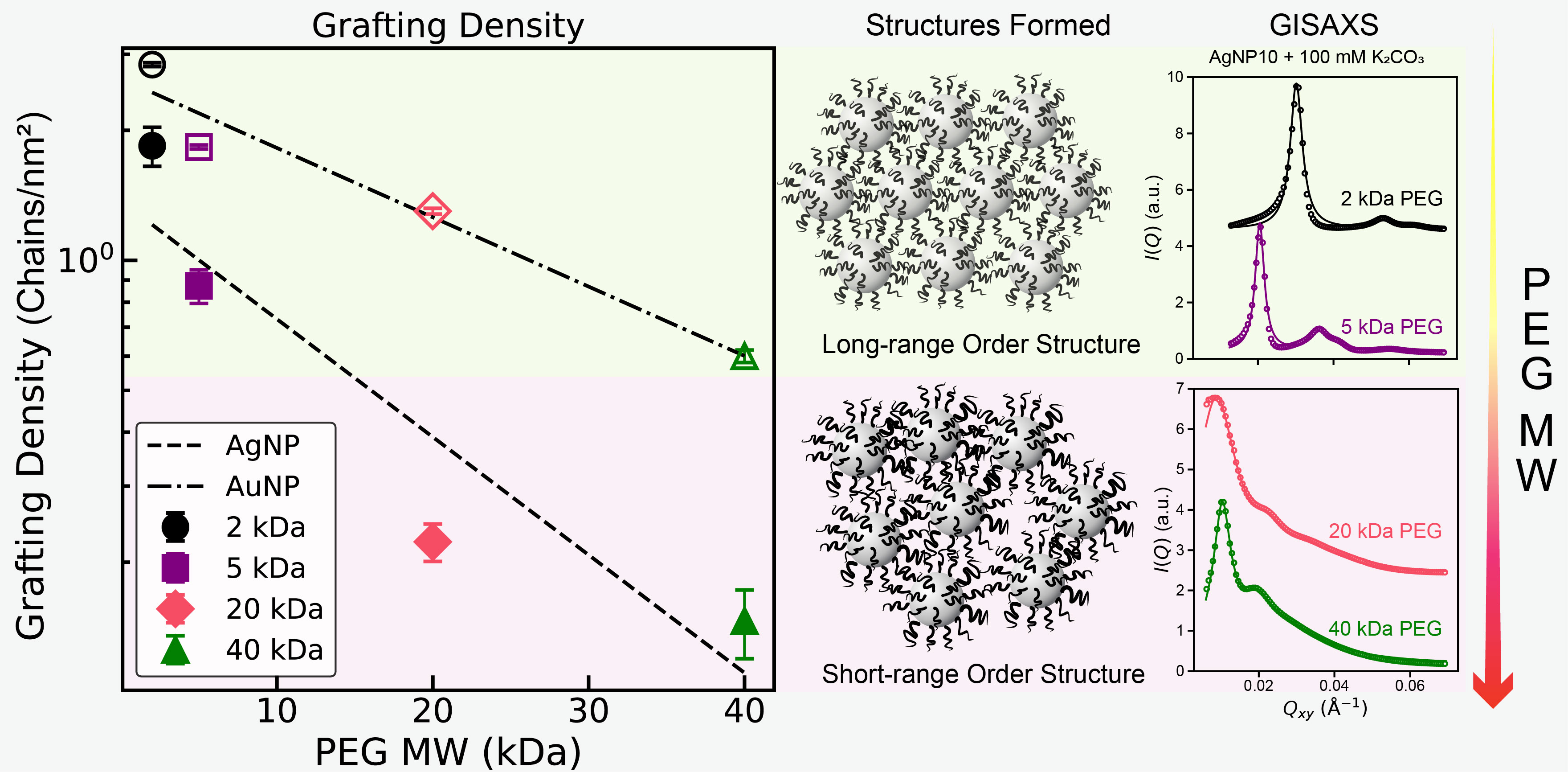}
    \label{fig:TOC}
\end{figure}

\section{Introduction}

The assembly of colloidal nanoparticles (NPs) into well-ordered superstructures is essential in fabricating advanced materials and devices with novel properties.\cite{young2014using,huh2020exploiting,bassani2024nanocrystal,travesset2024nanocrystal} The metal composition of colloidal NPs significantly influences their optical, electronic, and catalytic behaviors, making them vital for applications ranging from optoelectronics to catalysis. The ability to fine-tune the electromagnetic responses of NP lattices based on their composition provides versatility in designing functional materials and devices.\cite{polshettiwar2009self,zaera2013nanostructured,nie2010properties,yadav2021state} To accomplish the goal of creating functional devices based on the assembly of NPs, a variety of techniques have been developed.\cite{grzelczak2019stimuli,grzelczak2010directed,talapin2010prospects} Among these methods, the solvent evaporation technique stands out for its effectiveness in creating ordered structures, particularly when leveraging alkane-thiols with a high grafting density.\cite{murray1995self,liu2002evaporation,boles2016self,ye2024nanoparticle} Another route involves the incorporation of biopolymers, such as proteins or DNA, as well as water-soluble synthetic polymers.\cite{macfarlane2011nanoparticle,macfarlane2020nanoparticle} This approach takes advantage of the unique properties of these biological and synthetic macromolecules that are responsive to various aqueous conditions (electrolytes, pH, and temperature).   

Robust methods for assembling metal nanoparticles, particularly gold nanoparticles (AuNPs), by functionalizing them with water-soluble polymers such as polyethylene glycol (PEG) and poly(N-isopropylacrylamide) (PNIPAM) have been established.\cite{zhang2017macroscopic,kim2020temperaturenanorods,nayak2023assembling,nayak2023tuning} These polymers facilitate phase separation under specific electrolyte conditions or thermal stimuli, guiding the self-assembly of NPs into well-ordered two-dimensional (2D) at the vapor/liquid interfaces and three-dimensional (3D) structures in bulk solutions.\cite{zhang2017macroscopic,kim2020temperaturenanorods,nayak2023assembling,nayak2023tuning} This assembly technique is central for developing materials with enhanced optical, catalytic, and electronic properties for use in next-generation devices.

Building on geometric principles that favor binary lattice formation methods to create binary superlattices with well-defined stoichiometries such as AB, AB$_2$, A$_2$B, and AB$_3$ have been developed. \cite{kim2022binary,nayak2023ionic,kunzle2016binary} These superlattices offer superior tunability and unique properties that cannot be achieved with single-component assemblies. By grafting AuNPs with PEG chains of varying lengths or modifying their surface charge, the interparticle interactions can be precisely controlled, creating diverse superstructures.

It has been documented that silver nanoparticles (AgNPs) exhibit lower grafting densities with thiolated PEG compared to AuNPs due to the weaker or selective affinity of Ag for thiol groups. \cite{kim2022lamellar,liu2018controlled,marchioni2020thiolate,stewart2012controlling} This variation in grafting density may lead to assembling more complex binary superstructures, as the differences can be leveraged to drive selective inter-particle interactions, including directional binding. Such interactions may lead to novel properties in composite binary systems. Additionally, previous studies have shown that the core size of AgNPs affects grafting density due to surface curvature. Larger AgNPs, with a core size of 20 nm, tend to form less ordered structures at the vapor/liquid interface.\cite{kim2022lamellar} In this study, we employ smaller AgNPs with a 10 nm core to reveal distinct interfacial arrangements between the two core sizes. Additionally, we investigate the effect of shorter PEG chain lengths compared to those explored in previous studies.

Here, we present our findings on the effect of grafting density on the assembly of PEG-AgNPs, as well as binary systems composed of PEG-AuNPs and PEG-AgNPs. Using surface-sensitive synchrotron X-ray diffraction techniques, we characterize the structures of films formed from these grafted NPs at the vapor-liquid interface. By systematically adjusting suspension conditions, particularly pH and salinity, we demonstrate how the assembly and ordering of PEG-AgNPs and PEG-AuNPs can be controlled to create stable, well-ordered superstructures. These findings expand on previous work and offer a pathway for engineering complex nanoparticle assemblies with tailored properties, paving the way for potential applications in advanced devices.

\section{Methods}
\subsection{Preparation of Materials}
\begin{figure}[!hbt]
    \centering
    \includegraphics[width=1\linewidth]{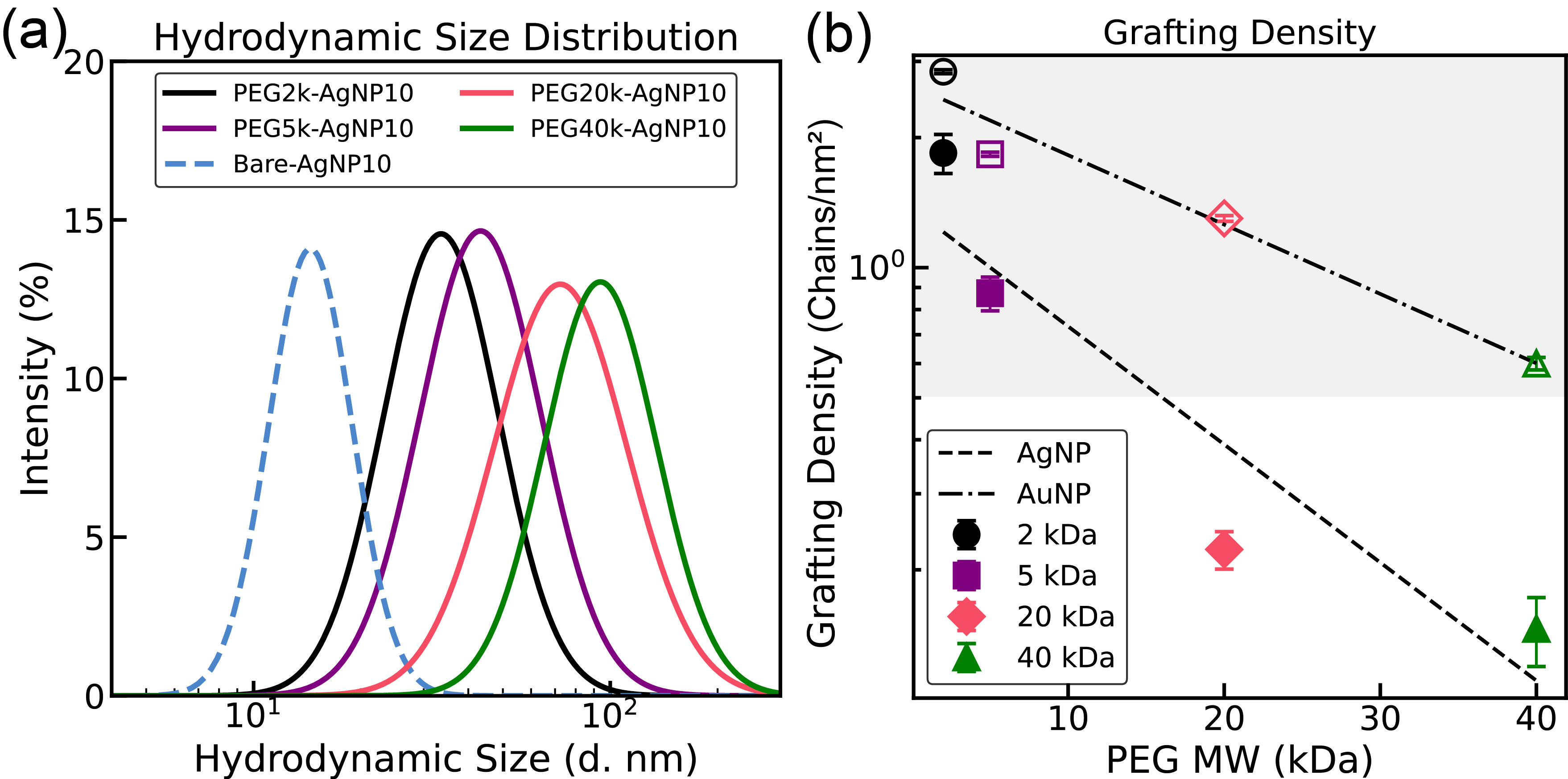}
    \caption{\scriptsize (a) DLS intensity percentage versus hydrodynamic size distribution for aqueous suspensions of bare AgNPs with a nominal size of approximately 10 nm (dashed line) and PEG-grafted AgNPs (solid lines) at room temperature. The peak positions in the DLS measurements shift to larger hydrodynamic sizes as the MW of the PEG increases, indicating successful grafting of PEG to the AgNP surfaces. (b) Grafting density of PEG-grafted AgNPs and AuNPs as a function of the molecular weight of the grafted PEG obtained from TGA measurements (see SI for details). The grafting density decreases as the PEG chain length (MW) increases, which is consistent with the behavior expected due to polymer folding and fanning out on the nanoparticle surfaces. The grafting-density vs. PEG MW plot is divided into two regions: a shaded area representing higher densities, which generally produce highly ordered hexagonal structures, and a lower-density region, where the structures are poorly ordered. A power law fit (as a guide to the eye) is applied to the data, represented by dotted lines for the AgNPs and dot-dashed lines for the AuNPs.(refer to Figure \ref{fig:TGA}).}
    \label{fig:DLS_TGA_main}
\end{figure}

Citrate-stabilized AgNPs with a nominal core diameter of $\sim$10 nm were purchased from Nanocomposix. Citrate-stabilized AuNPs with a nominal core diameter of $\sim$ 5 and 10 nm were purchased from Ted Pella Inc. Thiolated poly(ethylene glycol) (HS-PEG) with molecular weights (MW) of 2, 5, 20, and 40 kDa were purchased from Creative PEGWorks. Poly(acrylic acid) (PAA) with a MW of 2 kDa was purchased from Sigma-Aldrich. Hydrochloric acid (HCl) and potassium carbonate (\ch{K2CO3}) were purchased from Fisher Scientific and used without further purification. Milli-Q water (resistivity 18.2 M$\Omega$·cm at 25 \textdegree C) was used in all experiments. AgNPs and AuNPs were functionalized with HS-PEG by the ligand exchange method described elsewhere \cite{kim2022lamellar,nayak2023tuning}. Briefly, PEG ligands were dissolved in Milli-Q water and thoroughly mixed. An excess amount of PEG suspension was mixed with NPs, using specific ratios: 1 part of 10 nm AuNPs/AgNPs to 6000 parts of PEG, and 1 part of 5 nm AuNP to 1500 parts PEG. The resulting mixture was incubated for at least 24 hours with continuous mixing ($\sim$ 35 RPM) using a Roto-Shake Genie (Scientific Industries, NY, USA). The unbound PEG of the resulting PEG-grafted AuNPs was removed by centrifugation three times at a relative centrifugal force (RCF) of 21000 g for 90 minutes.  In this study, the term PEG-AuNPs represents PEG-grafted AuNPs, whereas PEG-AgNPs represent PEG-grafted AgNPs. PEG\textit{x}-AgNP\textit{y} represents AgNPs with a core diameter \textit{y} (i.e., $y=10$ nm) grafted with PEG chains of molecular weight \textit{x} (i.e., 2, 5, 20, or 40 kDa). To prepare PEG-AuNP / PEG-AgNP for experiments, the concentration of the suspension of NP was determined using ultraviolet-visible (UV-Vis) spectroscopy (NanoDrop One Microvolume, Thermo Fisher Scientific) and then adjusted to $\sim$12 nM for PEG-AgNP10, $\sim$ 20 nM for PEG-AuNP10 and $\sim$60 nM for PEG-AuNP5.
Subsequently, to confirm successful grafting, the hydrodynamic diameter ($D_{\rm H}$) of the grafted nanoparticles was determined using dynamic light scattering (DLS) with a NanoZS90 and its associated software: Zetasizer (Malvern, United Kingdom). Figure \ref{fig:DLS_TGA_main} shows the $D_{\rm H}$ distribution of these nanoparticles, indicating a systematic increase in $D_{\rm H}$ for the PEG-NPs with MW of PEG. This confirms the successful grafting of PEG to the NPs. To determine the average grafting density of PEG on each NP surface, thermogravimetric analysis (TGA) was conducted using a Netzsch STA449 and its associated software, Proteus. The grafted NPs were degraded in a temperature range of 100 to 600 $^{\circ}$C under argon gas. PEG grafted on AuNP/AgNP surfaces was found to thermally degrade between 350 to 450 $^{\circ}$C (as shown in Figure \ref{fig:TGA}). The grafting density ($\sigma$) was calculated using the method described in Ref.~\citenum{kim2023two} and is summarized in Figure \ref{fig:DLS_TGA_main}. Prior to X-ray experiments, a sufficient amount of stock solutions of \ch{K2CO3}, HCl, and PAA were prepared at high concentrations. These stock solutions were added in small amounts to the NP suspensions to achieve the desired electrolyte concentrations for the X-ray diffraction experiments.

For the X-ray diffraction experiments, single NP systems were directly loaded into a stainless steel trough. For binary systems, the NP suspensions were mixed in a desired molar ratio and incubated for 15 minutes before loading onto the trough. The trough containing the sample was placed in an enclosed chamber purged with water-saturated helium.\cite{pershan2012liquid} A calculated amount of stock electrolyte solutions was added incrementally to the same suspension of the mixture to achieve a sequence of target electrolyte concentration values, as reported in Table \ref{table:PEG_conditions}.
\begin{table*}[!htb]
\caption{Summary of hydrodynamic diameters from dynamic light scattering (DLS) and grafting densities from thermogravimetric analysis (TGA). The data presented in this table were used to generate Figure \ref{fig:DLS_TGA_main}}
\begin{threeparttable}
\normalsize
\begin{tabular}[htp]{ccc|ccc}
    \toprule \toprule
    \multicolumn{3}{c|}{DLS} & \multicolumn{3}{c}{TGA}\\
    \midrule
     {NP} & { MW of grafted} & {Hydrodynamic}& {Initial weight}  & {Changed } & {Grafting density}  \\
    {type} & {PEG (kDa) } & {size (nm)\tnote{1}}& {(mg)} & {weight\tnote{3}} &{(chains/nm$^2$)}\\
    \midrule
    \multirow{5}{*}{AgNP10} & n.a.\tnote{2} & 16.1$\pm$0.5 & - & - & - \\
    & 2 & 41.2$\pm$0.8 & 1.84 & 23\% & $1.85\pm0.19$ \\
    & 5 & 54.4$\pm$1.0 & 2.30 & 28\% & $0.87\pm0.08$ \\
    & 20 & 95.5$\pm$1.4 & 1.18 & 26\% & $0.22\pm0.02$ \\
    & 40 & 114.4$\pm$1.4 & 0.94 & 30\% & $0.14\pm0.15$ \\
    \bottomrule \bottomrule
\end{tabular}
\begin{tablenotes}
    \item[1] {\scriptsize Only the modal size on the distribution profile is reported.}
    \item[2] {\scriptsize The bare surface AgNPs, stabilized with citrate ligands.}
    \item[3] {\scriptsize The changed weight was calculated from Fig. \ref{fig:TGA} and is relative change with respect to initial weight.}
\end{tablenotes}
\end{threeparttable}
\label{tbl:dls-tga}
\end{table*}

\subsection{X-ray Diffraction Setup}
Synchrotron-based \textit{in-situ} liquid surface X-ray scattering experiments were performed at NSF’s ChemMatCARS Sector 15, Advanced Photon Source (APS), Argonne National Laboratory, and at SMI beamline open platform liquid surfaces (OPLS) end station, National Synchrotron Light Source II (NSLS-II), Brookhaven National Laboratory with an incident X-ray energy 10 and 9.7 keV, respectively. Pilatus area detectors were used to record scattered X-ray beams.

Liquid-surface specular X-ray reflectivity (XRR) was conducted to determine the electron density (ED) profile, $\rho(z)$, along the axis normal to (the $z$ direction, orthogonal coordinates) the vapor-liquid interface. Grazing incidence small-angle X-ray scattering (GISAXS) was used to analyze the nanoparticle assembly's in-plane ($x,y$ directions) arrangement at the aqueous surface. In both cases, the collimated monochromatic X-rays of wavevector $\vec{k}_{\rm i}$ are incident on the liquid surface at grazing incident angles, and the scattered X-rays with wavevector $\vec{k}_{\rm f}$ are recorded with an area detector.
The scattering vector, $\vec{Q}$, is equal to $\vec{k}_{\rm f}-\vec{k}_{\rm i}$. $Q_{z}$ and $Q_{xy}$ denote the vertical and horizontal components of $\vec{Q}$, respectively. 
For XRR, the reflectivity $R(Q_z)$ profiles, normalized to the Fresnel reflectivity $R_{\rm F}$, are analyzed in terms of electron density (ED) profiles via Paratt's recursive method.\cite{Nielsen2011} The GISAXS intensities are mapped as a function of $(Q_{xy}, Q_{z})$, and characteristic rod-like scattering patterns are typical of two-dimensional ordered superlattices. Linecut intensity profiles, obtained by integrating intensities over a narrow $Q_z$ range and denoted as $I(Q_{xy})$ correspond to 2D in-plane arrangements and, for crystalline systems, are labeled with 2D Miller indices $(h, k)$. X-ray data collection and processing were conducted on-site according to each beamline's protocols. Details about the instrumental setup and measurement protocols are detailed in previous reports.\cite{kim2022binary,kim2022lamellar,zhang2017macroscopic,nayak2023assembling}

\section{Results and Discussions}
AgNPs pose unique challenges when it comes to grafting SH-PEG due to the lower affinity of silver for thiol groups compared to AuNPs. The weaker Ag–S bond makes it more difficult to achieve stable grafting, unlike the stronger Au–S bond observed in AuNP systems.\cite{liu2018controlled,marchioni2020thiolate,stewart2012controlling} To confirm successful PEG grafting onto AgNPs, we first conducted DLS measurements. These measurements revealed an increase in the $D_{\rm H}$ of PEG-AgNPs compared to bare AgNPs, as shown in Figure \ref{fig:DLS_TGA_main}, indicating successful surface functionalization.

We measured the grafting density of PEG-AgNPs with varying MWs. As anticipated, the grafting density decreased with increasing PEG MW, likely due to polymer folding on the nanoparticle surface (see Table \ref{tbl:dls-tga} and Figure \ref{fig:DLS_TGA_main}). Figure \ref{fig:DLS_TGA_main} further confirms that the grafting density of PEG on AgNPs is consistently lower than that on AuNPs. This is consistent with previous results that show PEG grafting AgNPs (20 nm core size) is less efficient when compared to AuNPs. \cite{kim2022lamellar}

\begin{figure}[!ht]
    \centering
    \includegraphics[width=1\linewidth]{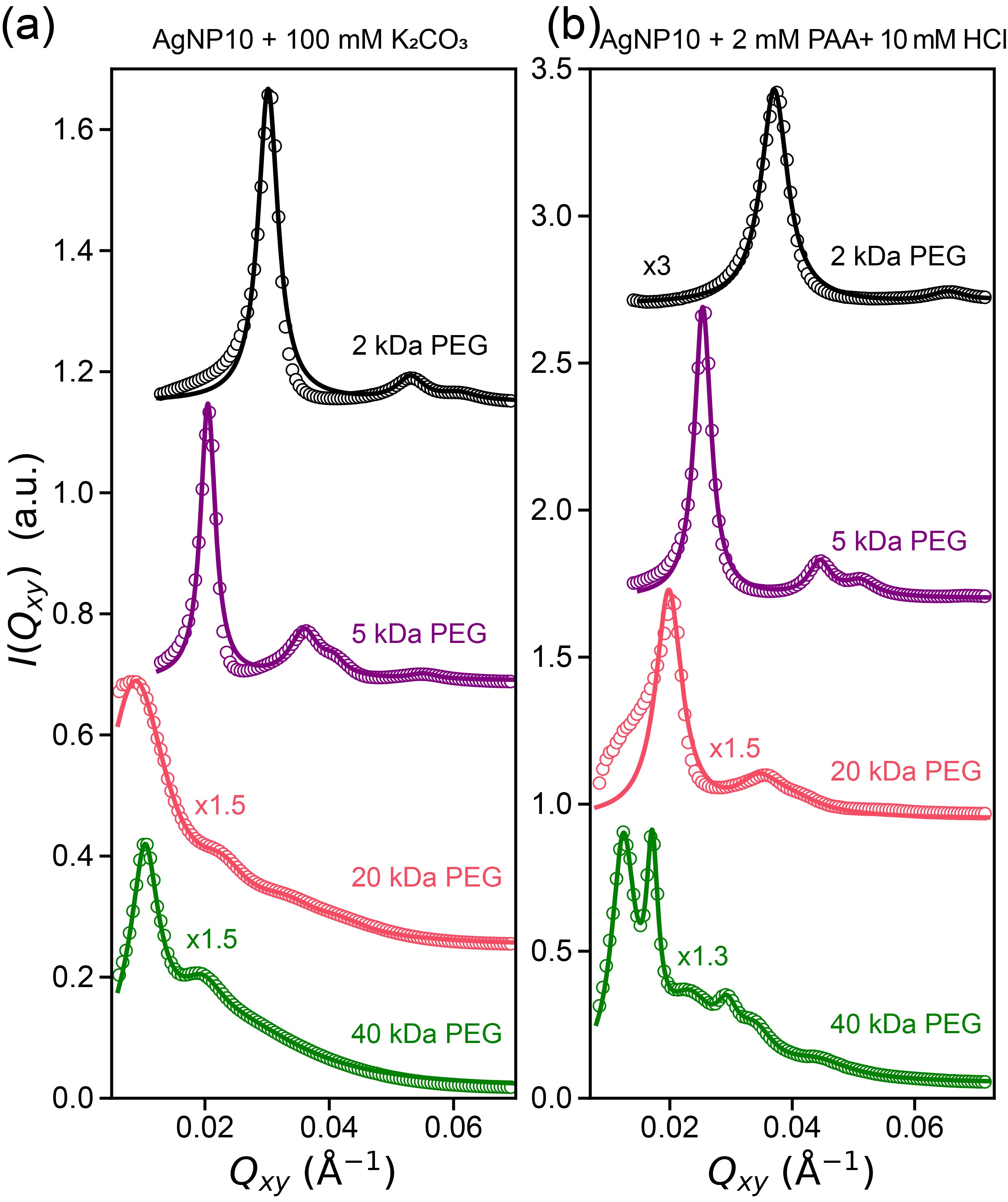}
    \caption{\scriptsize GISAXS linecut profiles displaying intensity ($I(Q_{xy})$) versus in-plane momentum transfer ($Q_{xy}$) for PEG-AgNPs with various MW in suspensions as indicated. (a) in 100 mM \ch{K2CO3}, (b) in 2 mM PAA in the presence of 10 mM HCl. The solid lines represent the best-fit Lorentzian profiles to the experimental data. The plots are vertically shifted for clarity, and in some cases, the intensity of the plots is multiplied by a constant, as indicated, to enhance visibility.}
    \label{fig:gsx_main}
\end{figure}

We begin by examining the influence of different electrolytes (either \ch{K2CO3} or PAA with HCl) on the structural ordering of AgNPs grafted with PEG. The goal is to better understand how the PEG chain length (MW) affects the in-plane ordering of PEG-AgNPs under these suspension conditions. Figure \ref{fig:gsx_main} presents GISAXS line-cut profiles, showing the intensity of scattering ($I(Q_{xy})$) as a function of in-plane momentum transfer ($Q_{xy}$) for PEG-AgNPs of different MWs. Panel (a) shows the results for suspensions in 100 mM \ch{K2CO3}, and panel (b) for 2 mM PAA with 10 mM HCl. \ch{K2CO3} was chosen for its significant effect on phase separation of PEG in aqueous solutions \cite{huddleston2003phase}, while PAA under acidic conditions was used to promote interpolymer complexation (IPC) by hydrogen bonding between PAA and PEG.\cite{zhang2017macroscopic,Nayak2019IPC,kim2021effectsIPC} The solid lines in the figures represent best-fit Lorentzian-shaped peaks, which provide insight into the in-plane ordering of the nanoparticles. The parameters extracted from these analyses and associated structural parameters are listed in Table \ref{table:PEG_conditions}.

In Figure \ref{fig:gsx_main} (a), for AgNP suspensions in 100 mM \ch{K2CO3} we observe that PEG-AgNPs with MWs of 2 and 5 kDa form a clear hexagonal structure, with sharp diffraction peaks. This indicates a regular spacing between particles, driven by strong van der Waals interactions influenced by the short PEG chains. As the PEG MW increases beyond 5 kDa, the profiles exhibit a transition to a less ordered hexagonal structure, likely due to increased steric hindrance or lower grafting density of the longer polymer chains.

In Figure \ref{fig:gsx_main} (b), for suspensions in 2 mM PAA with 10 mM HCl, we observe that PEG chains with MWs between 2 and 20 kDa maintain a 2D hexagonal arrangement, similar to the \ch{K2CO3} system. However, for PEG chains with a molecular weight of 40 kDa, the profiles indicate the formation of two distinct structural phases, which are discussed in more detail below. This induced phase separation is likely driven by complex interparticle interactions and low or irregular grafting density from long-chain PEG, balancing steric effects from the PEG with electrostatic interactions from PAA/PEG and HCl.

A key trend in both electrolyte conditions is that the primary diffraction peaks shift to lower $Q_{xy}$ values as the PEG MW increases. This shift corresponds to an expansion of the lattice unit cell, as longer PEG chains increase the spacing between nanoparticles (see Table \ref{table:PEG_conditions}). This shows that the length of the PEG chain directly controls the assembly of the nanoparticles by adjusting the nearest-neighbor distance ($d_{\rm NN}$), which is consistent with the literature reported.\cite{kim2021effectchain} The length of the PEG chain and the electrolyte environment play crucial roles in the structural organization of AgNPs at the vapor-liquid interface. Shorter PEG chains lead to well-ordered hexagonal structures, while longer chains cause disorder and phase separation.

\begin{figure}[!hbt]
    \centering
    \includegraphics[width=1\linewidth]{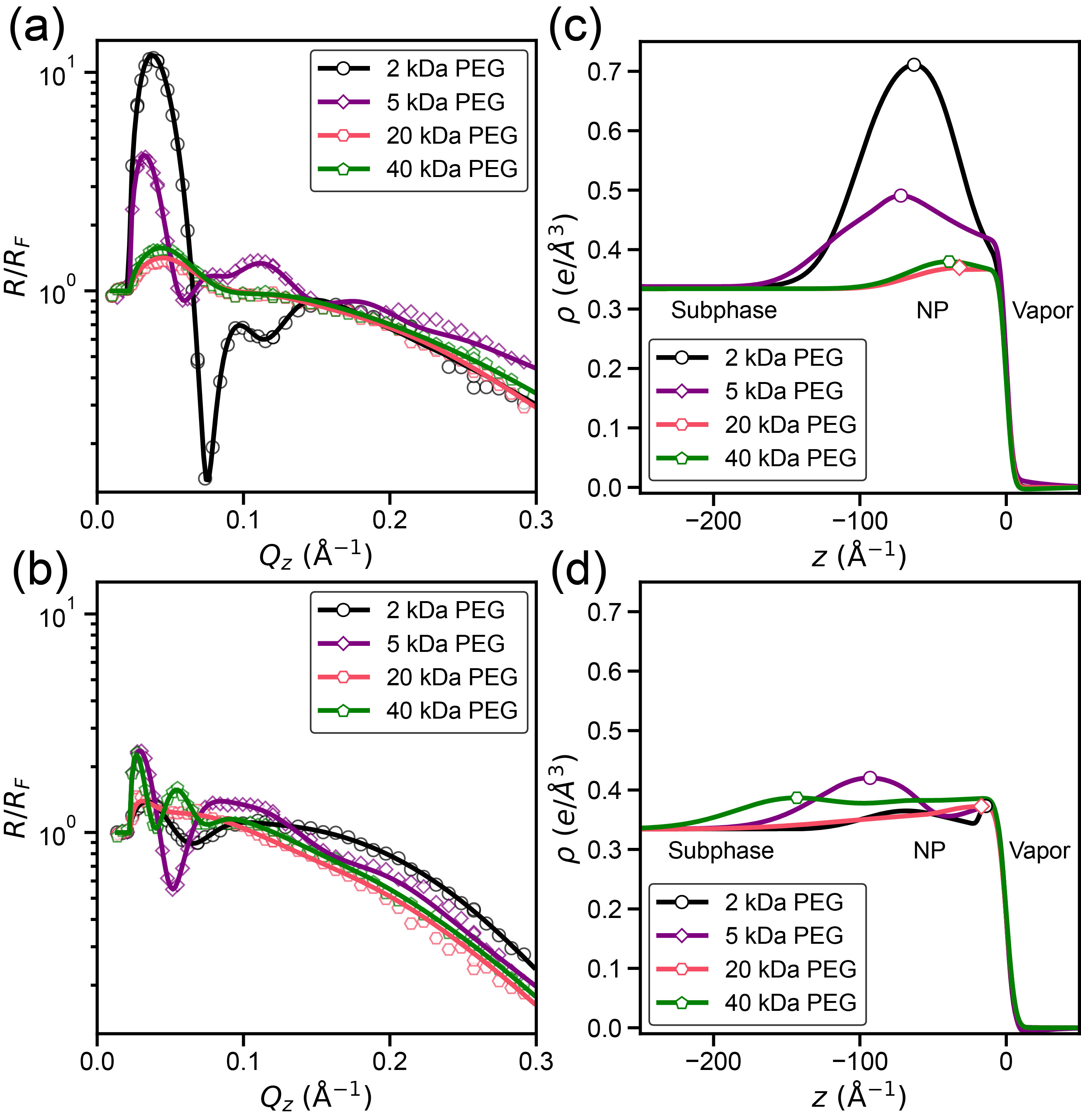}
    \caption{\scriptsize Normalized X-ray reflectivity profiles \((R/R_{\text{F}})\) for the same films shown in Figure \ref{fig:gsx_main}, representing: (a) PEG-AgNPs in a 100 mM \ch{K2CO3} solution, and (b) PEG-AgNPs in a 2 mM PAA solution with the addition of 10 mM HCl, with PEG MWs as indicated. The ED profiles that are used to fit the XRR data (solid lines) in (a) and (b) are presented in (c) and (d).}

    \label{fig:ref_main}
\end{figure}

\subsection{XRR and ED Analysis of PEG-AgNP Film}

XRR measurements were conducted on the same PEG-AgNP films to examine the film profiles formed at the vapor-liquid interface. These measurements serve to corroborate the GISAXS results discussed above, providing additional information on the particle distribution and structural arrangement at the vapor-liquid interface. Figure \ref{fig:ref_main} presents normalized XRR scans \((R/R_{\text{F}})\) for PEG-AgNP films, with (a) showing films in 100 mM \ch{K2CO3}, and (b) depicting films in a 2 PAA solution with the addition of 10 mM HCl. The MW of PEG used in each case is indicated. The corresponding ED profiles derived from the fitting of the XRR data are shown in (c) and (d), respectively.

Qualitatively, the normalized XRR  for films with short-chain PEG exhibits significantly higher reflectivity compared to those with longer PEG chains. This enhanced reflectivity can be intuitively attributed to a greater surface particle density at the vapor-liquid interface, as shorter PEG chains allow for denser packing of NPs at the surface. Indeed, the strongest enhancement is seen for PEG-AgNPs grafted with the shortest chain (2 kDa), confirming that the surface density is highest for these systems.

\begin{figure*}[!hbt]
    \centering
    \includegraphics[width=1\linewidth]{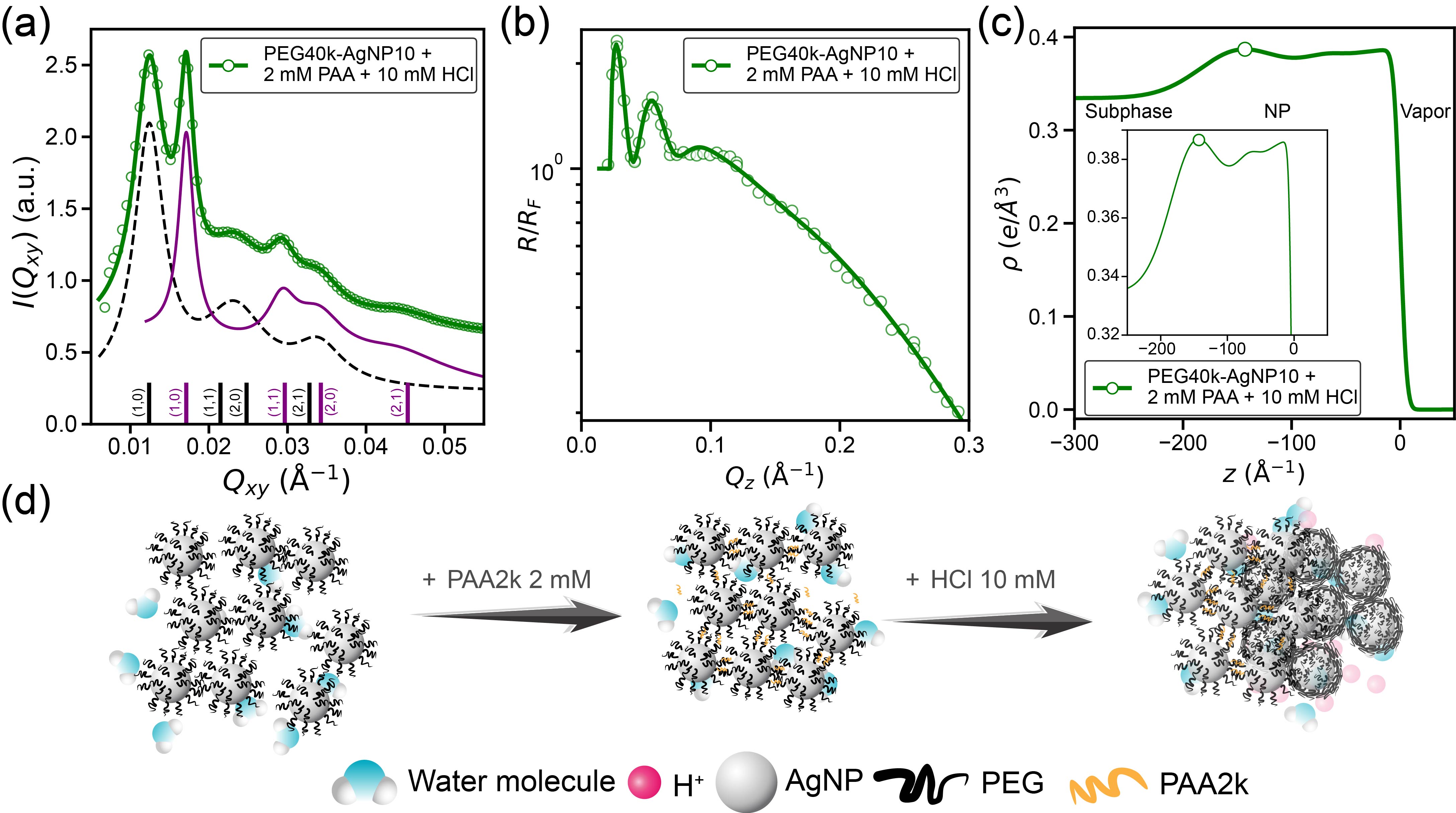}
    \caption{\scriptsize (a) GISAXS line-cut profiles showing intensity \((I(Q_{xy}))\) versus in-plane momentum transfer \((Q_{xy})\) for PEG40k-AgNPs in a suspension containing 2 mM PAA and 10 mM HCl. The solid lines represent the best-fit Lorentzian profiles to the experimental data. The diffraction pattern analysis suggests the formation of two distinct phases, represented by the solid and dashed lines, each corresponding to a hexagonal structure. One phase exhibits a higher-quality crystal structure with a smaller lattice constant compared to the other. (b) Normalized X-ray reflectivity profiles \((R/R_{\text{F}})\) for the same film depicted in (a), with the best-fit solid line derived from the electron density (ED) profiles shown in (c). The inset illustrates the enhanced ED profile relative to the subphase, indicating that the film thickness is greater than the diameter of a single PEG-AgNP. (d) Schematic depiction of the two-dimensional structures formed at the vapor-liquid interface, where the two crystalline phases are likely induced by the presence of PAA in the suspension.}
    \label{fig:40k_main}
\end{figure*}

Quantitatively, this trend is supported by fitting the XRR data to the corresponding ED profiles shown in Figures \ref{fig:ref_main} (c) and (d). The bell-shaped peaks in these profiles, located near the vapor-liquid interface, are characteristic of the metallic AgNPs, which have ED higher than that of the surrounding water or grafted PEG. The width of these peaks provides an estimate of the average particle diameter, while the integral of the peak offers a measure of the particle density at the vapor-liquid interface (see Table \ref{table:PEG_conditions}). As expected, the ED profiles confirm that films formed with short-chain PEG exhibit a higher particle density at the vapor-liquid interface, consistent with the structural information obtained from GISAXS data, where the unit cell dimensions also suggest denser nanoparticle packing for shorter PEG chains. The consistency between XRR and GISAXS results indicates the presence of uniformly ordered structures at the interface, with no evidence of isolated patches of ordered NPs on the surface. \cite{zhang2017macroscopic}

In the \ch{K2CO3} solution, the AgNP films form a well-defined single-layer structure at the vapor-liquid interface. In contrast, when PAA and HCl are added to the system, the ED profiles suggest the formation of incomplete bilayer structures for PEG with MWs of 20 and 40 kDa. However, these bilayer structures exhibit a lower surface density compared to the single-layer formations observed with shorter PEG chains. This transition to a bilayer structure may be attributed to the lower grafting density for longer PEG chains facilitating polymer interdigitation, leading to a looser in-plane packing.

The XRR data, along with the electron density profiles, offer complementary insights into how PEG chain length and the electrolyte environment influence the arrangement of AgNPs at the vapor-liquid interface. Short-chain PEG leads to higher surface densities and more uniform monolayer structures, while longer PEG chains induce the formation of more complex but less dense bilayer structures. This is consistent with previous results of PEG-AuNPs and PNIPAM-AuNPs under similar conditions. \cite{zhang2017macroscopic,kim2021effectchain, nayak2023assembling}

\subsection{Quasi-bilayered ordering from PEG40k-AgNPs}
\begin{figure*}[!hbt]
    \centering
    \includegraphics[width=1\linewidth]{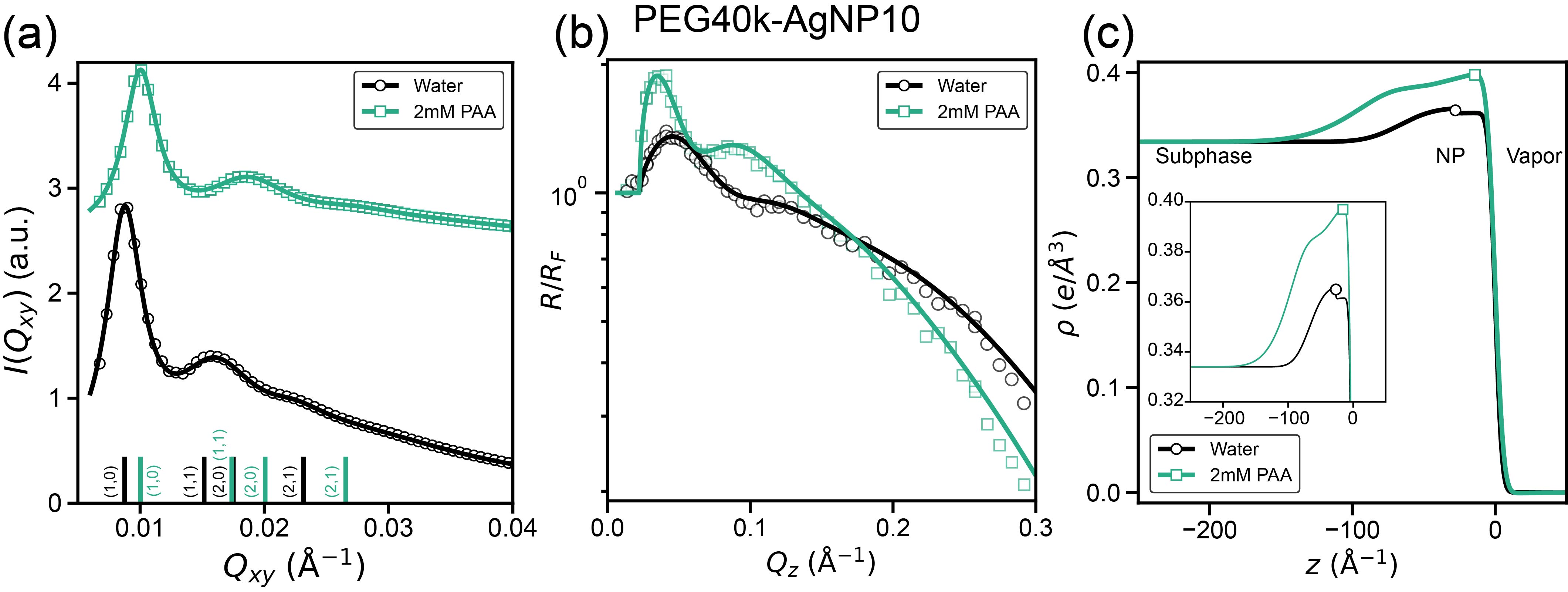}
    \caption{\scriptsize Control expiements to justify the interpretation of Figure \ref{fig:40k_main}. (a) GISAXS line-cut profiles showing intensity \((I(Q_{xy}))\) versus in-plane momentum transfer \((Q_{xy})\) for PEG40k-AgNPs in a suspension containing water ($\bigcirc$) and 2 mM PAA ($\square$) (without HCl). The solid line is the best-fit Lorentzian profiles for the diffraction data, assuming a short-range hexagonal structure. The second and third harmonic peaks are broadened due to inplane irregularities. (b) Normalized XRR profiles \((R/R_{\text{F}})\) for the same film depicted in (a), with the best-fit (solid line) derived from the ED profiles shown in (c). The inset illustrates the enhanced ED profile relative to the subphase, indicating that the film thickness consists of a single layer of PEG-AgNPs.}
    \label{fig:40k_addl}
\end{figure*}

We now proceed to a more detailed analysis of PEG40k-AgNP immersed in a solution of 2 mM PAA and 10 mM HCl, as described earlier, where the system exhibits ordering into two distinct phases. Figure \ref{fig:40k_main} (a) shows the GISAXS line-cut profiles of the intensity \((I(Q_{xy}))\) versus in-plane momentum transfer \((Q_{xy})\) for PEG40k-AgNPs in this suspension. The solid lines represent the best-fit Lorentzian profiles derived from the experimental data. From the diffraction pattern, we identify two distinct hexagonal phases: one with a smaller unit cell and better crystallinity, represented by the solid line, and another with a larger unit cell, represented by the dashed line. These two phases likely correspond to structural differences within the presumptive quasi-bilayer of the film. The phase with the smaller lattice constant likely arises from nanoparticles with collapsed PEG chains, while the phase with the larger lattice constant is consistent with an expanded PEG configuration, leading to greater $d_{\rm NN}$. Figure \ref{fig:40k_main} (b) displays the normalized X-ray reflectivity profiles \((R/R_{\text{F}})\) for the same film shown in panel (a). The best-fit solid line is derived from the ED profiles shown in Figure \ref{fig:40k_main} (c). The ED profile inset illustrates a distinct enhancement at the vapor-liquid interface relative to the subphase, indicating that the film thickness exceeds the diameter of a single PEG-AgNP ($\sim 10$ nm). This suggests that the film consists of two (likely incomplete) ordered layers (a quasi-bilayer). The top layer, with a larger lattice constant, is associated with an expanded PEG structure, while the bottom layer, with a smaller lattice constant, consists of nanoparticles with collapsed PEG chains. We further support this interpretation by comparing these results to control experiments performed with PAA but without HCl, as shown in Figure \ref{fig:40k_addl}. In the absence of HCl, the GISAXS profile is very similar to the dashed-line profile from Figure \ref{fig:40k_main} (a), corresponding to the phase with a larger lattice constant. This similarity suggests that only a single ordered phase forms in the absence of HCl, with PEG chains adopting a more open configuration. The XRR data in Figure \ref{fig:40k_addl} (b) further corroborate this and its corresponding ED profile in Figure \ref{fig:40k_addl} (c), which shows that the film consists of a single layer at the vapor-liquid interface, with a thickness of approximately 10 nm, corresponding to the core diameter of a single AgNP.

Thus, the presence of HCl induces the formation of a second, more collapsed phase, likely due to changes in the electrostatic interactions, interdigitation processes, and PEG chain conformations. This dual-phase structure is a unique feature observed for PEG40k-AgNPs in the presence of 2 mM PAA and HCl, highlighting the role of the acidic environment in modulating nanoparticle ordering at the vapor-liquid interface.
\begin{figure}[!hbt]
    \centering
    \includegraphics[width=1\linewidth]{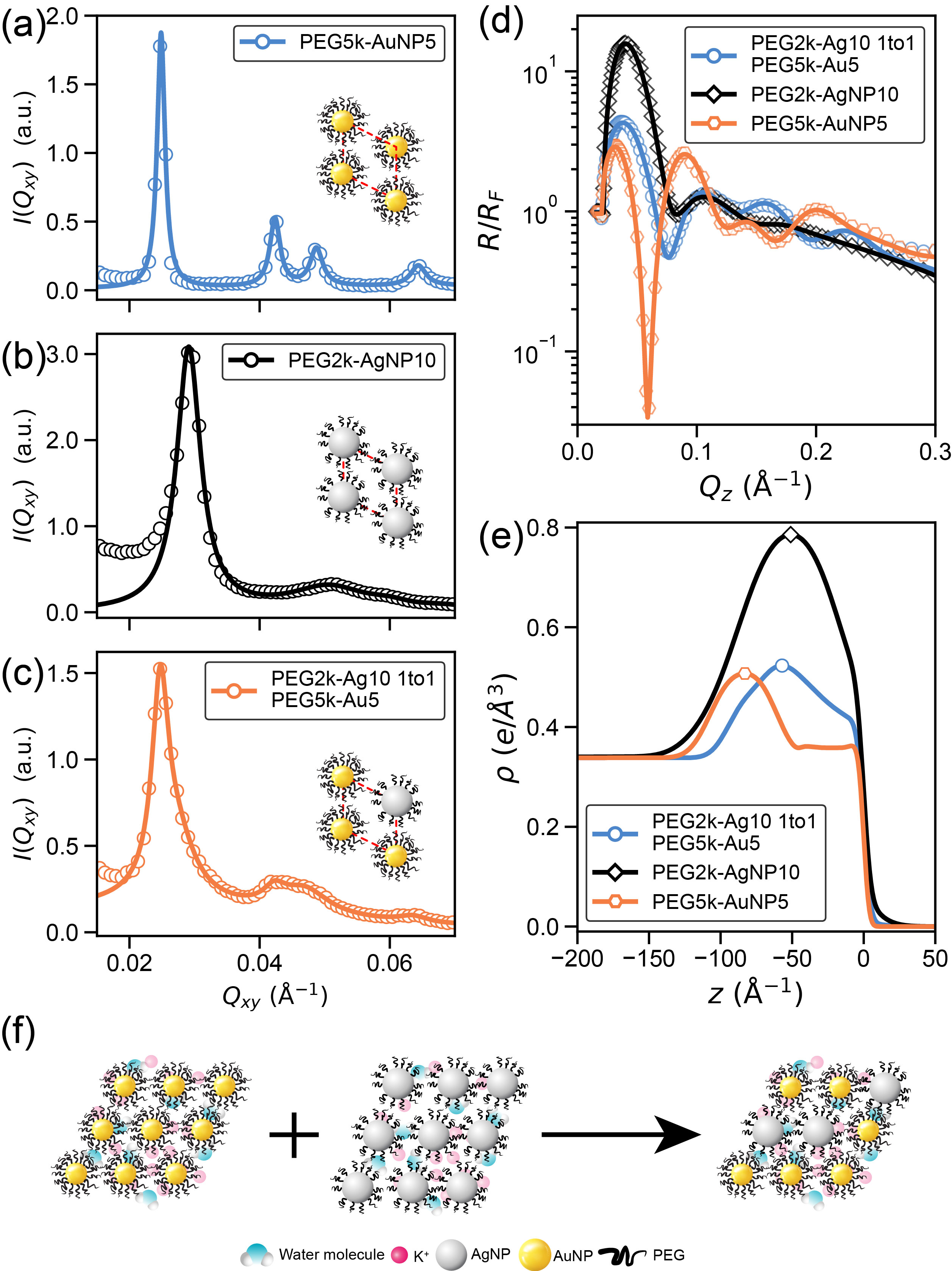}
    \caption{\scriptsize GISAXS line-cut profiles showing intensity \((I(Q_{xy}))\) versus in-plane momentum transfer \((Q_{xy})\) for (a) PEG5k-AuNP5, (b) PEG2k-Ag10, and (c) for a 1:1 binary mixture of PEG2k-AgNP10 and PEG5k-AuNP5 under 100 mM \ch{K2CO3} suspension conditions. The solid lines represent the best-fit Lorentzian profiles to the experimental data. (d) Normalized X-ray reflectivity profiles \((R/R_{\text{F}})\) for the same films depicted in (a-c), with the best-fit solid line obtained from the ED profiles shown in (e). (f) Schematic illustration of the lattices formed after mixing AuNPs and AgNPs.}
    \label{fig:compare}
\end{figure}


\subsection{Binary Systems of AuNPs and AgNPs}
The results above underscore the distinct behaviors of PEG-grafted AgNPs and AuNPs (the latter of which are briefly presented here for completeness). Extensive studies have previously reported on the influence of PEG MW on PEG-AuNPs under various electrolyte conditions, as well as on binary systems formed from PEG-AuNPs \cite{zhang2017macroscopic, kim2021effectchain, kim2022binary,nayak2023ionic}. Given the observed differences in grafting densities between AuNPs and AgNPs (Figure \ref{fig:DLS_TGA_main}), we aim to explore the assembly of binary systems combining both types of nanoparticles. Here, we present preliminary findings on molar mixing ratio conditions, which offer initial insights into the broader implications of such binary assemblies. 

Figure \ref{fig:compare} presents the GISAXS and XRR results for PEG5k-AuNP5 and PEG2k-AgNP10 as control experiments in a 1:1 mixing ratio (100 mM \ch{K2CO3}). Our results show that both PEG-AuNPs and PEG-AgNPs form well-ordered hexagonal structures. The XRR and ED profiles confirm the formation of a single layer at the vapor-liquid interface, while the ED profiles reveal a higher surface density for AgNPs, likely due to their shorter PEG chains. The GISAXS and XRR results for the mixture of these particles suggest that the binary system of PEG-AuNPs and PEG-AgNPs forms a hybrid structure, as both the XRR and GISAXS patterns differ from those of the individual components, as illustrated schematically in Figure \ref{fig:compare} (f). While the overall system remains ordered, it exhibits defects and is primarily composed of AuNPs with dispersed AgNPs, lacking a well-defined binary superstructure. This interpretation is further supported by the ED profiles in Figure \ref{fig:compare} (e), where AgNPs show a higher surface density compared to AuNPs. In the mixture, the ED profile shows a superposition of ED profiles of the individual components, although PEG-AgNPs are present at a lower concentration.

\begin{figure}[!hbt]
    \centering
    \includegraphics[width=1\linewidth]{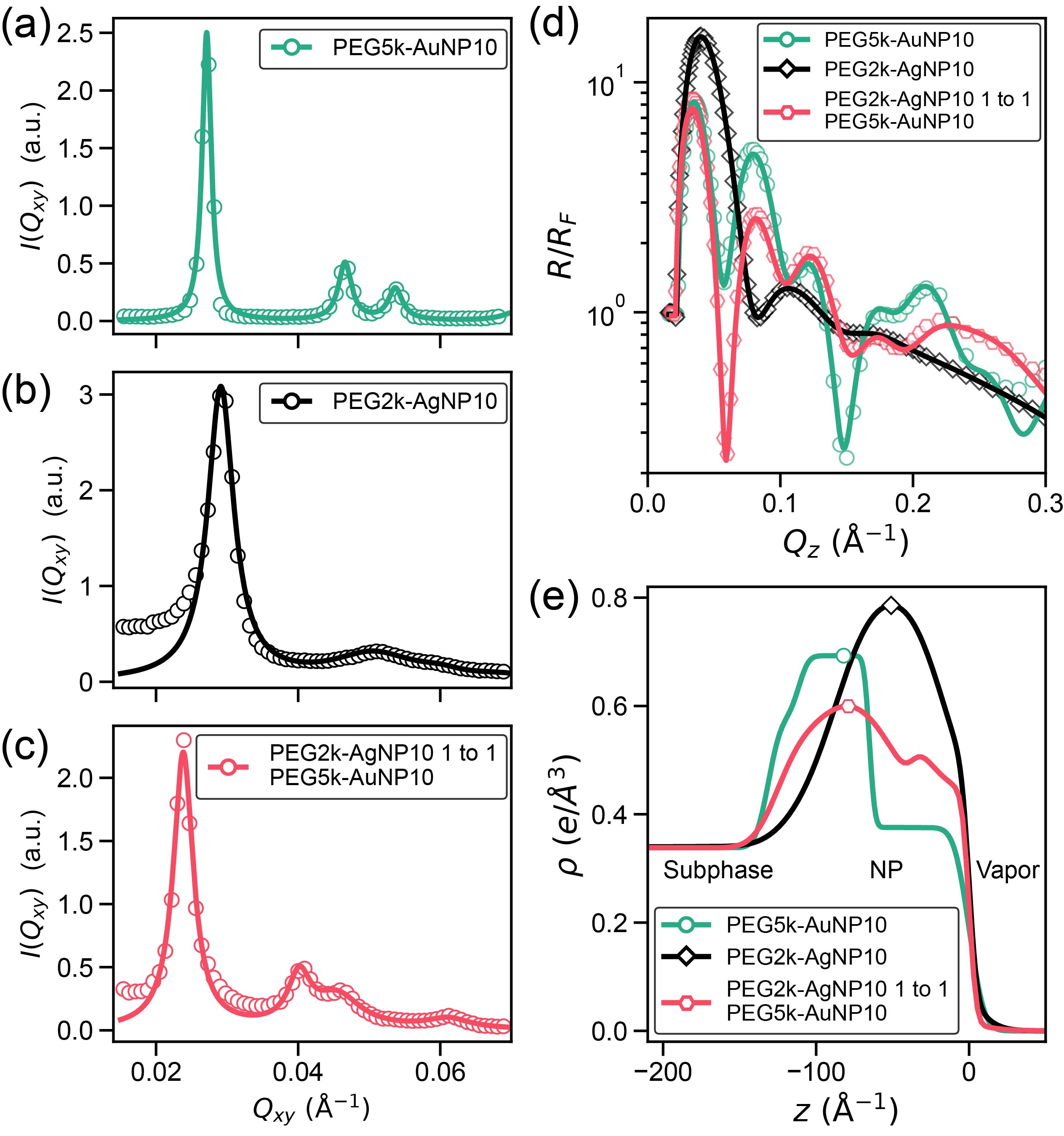}
    \caption{\scriptsize GISAXS line-cut profiles showing intensity \((I(Q_{xy}))\) versus in-plane momentum transfer \((Q_{xy})\) for (a) PEG5k-AuNP10 (b) PEG2k-AgNP10, and (c)  for 1:1 a binary mixture of PEG2k-AgNP10 and PEG5k-AuNP10, under 100 mM \ch{K2CO3} suspension conditions. The solid lines represent the best-fit Lorentzian profiles to the experimental data. (d) Normalized X-ray reflectivity profiles \((R/R_{\text{F}})\) for the same films depicted in (a-c), with the best-fit solid line obtained from the ED profiles shown in (e).}
    \label{fig:21050_compare}
\end{figure}

This trend, where AuNPs populate the surface more than AgNPs, is further confirmed in a 1:1 mixture of PEG5k-AuNP10 and PEG2k-AgNP10. Figure \ref{fig:21050_compare} presents the GISAXS and XRR results for PEG5k-AuNP10, with PEG2k-AgNP10 as control in a similar mixing ratio. The XRR and ED profiles continue to show that AgNPs, due to their shorter PEG chains, are enriched at the surface. As observed in the previous case, the binary system forms a hybrid structure. However, the diffraction pattern in this system shifts to lower $Q_{xy}$ values, suggesting an expanded lattice constant or short-range order due to the inhomogeneous formation of the binary lattice. This is further supported by the lattice constant values listed in the Table \ref{bianry_table}, indicating the formation of a more disordered superstructure. The ED profiles, as shown in Figure \ref{fig:compare} (e), reveal a higher surface density for AgNPs, with the mixture displaying a superposition of both nanoparticle types, albeit with PEG-AgNPs present at a lower concentration.

\begin{figure}[!hbt]
    \centering
    \includegraphics[width=1\linewidth]{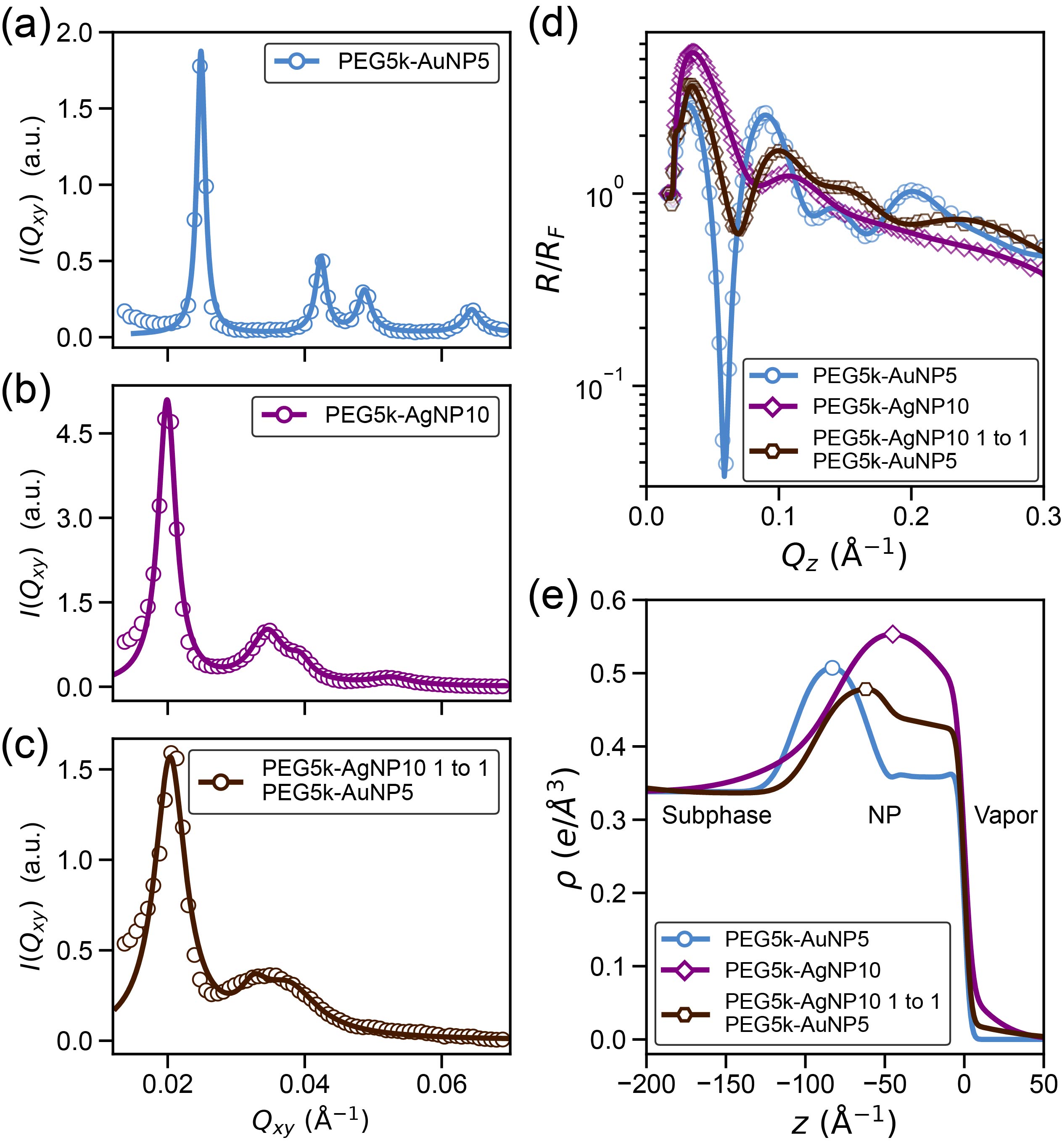}
    \caption{\scriptsize GISAXS line-cut profiles showing intensity \((I(Q_{xy}))\) versus in-plane momentum transfer \((Q_{xy})\) for (a) PEG5k-AuNP5, (b) PEG5k-Ag10, and (c) for a 1:1 binary mixture of PEG5k-AgNP10 and PEG5k-AuNP5 under 100 mM \ch{K2CO3} suspension conditions. The solid lines represent the best-fit Lorentzian profiles to the experimental data. (d) Normalized X-ray reflectivity profiles \((R/R_{\text{F}})\) for the same films depicted in (a-c), with the best-fit solid line obtained from the ED profiles shown in (e).}
    \label{fig:55510_compare}
\end{figure}

We further explore another combination of NPs, PEG5k-AuNP5 and PEG5k-AgNP10, by examining their structures and comparing them with the binary systems that are typically seen in PEG-AuNPs system.\cite{kim2022binary,nayak2023ionic} Figure \ref{fig:55510_compare} presents the GISAXS and XRR results for these nanoparticles, showing well-defined hexagonal lattices for both systems. The size of the core plays a key role in determining the lattice constants, with AuNP5 exhibiting a significantly smaller lattice constant than AgNP10, despite both being grafted with the same molecular weight PEG. The XRR and ED profiles, shown in Figures \ref{fig:55510_compare} (d) and (e), confirm these findings and highlight a higher surface density for AgNPs due to their larger core NP. In the mixed system, the structure appears to favor the AgNP lattice, as seen in the GISAXS data in Figure \ref{fig:55510_compare} (c), although both the GISAXS and XRR results suggest that AuNPs are incorporated into the AgNP matrix, disrupting the overall order of the superlattice. The binary system exhibits less order compared to the individual components, indicating a disruption of the lattice structure when the two types of nanoparticles are combined.

The investigations into binary systems of PEG-grafted AuNPs and AgNPs reveal distinct behaviors between the two nanoparticle types, influenced by differences in PEG chain length and nanoparticle core size. Both individual systems form well-ordered hexagonal structures, but in mixed systems, the binary assemblies exhibit hybrid structures with disrupted order and expanded lattice constants. In most cases, AuNPs dominate the surface arrangement, while AgNPs fuse into the system, contributing to the formation of a less ordered hybrid structure.

This investigation is preliminary, and further work is required to achieve more complex binary assemblies. Future studies could focus on varying the concentration of AgNPs in the mixture to promote superstructure formation, as seen in AuNP-only systems.\cite{kim2022binary} Additionally, grafting nanoparticles with charged terminal PEG chains could introduce ionic interactions, potentially forming ionic-like superlattices, as demonstrated in recent studies.\cite{nayak2023ionic} These approaches may refine the structural control of AuNP-AgNP binary systems and enhance their overall ordering. It is important to note that these results suggest a critical grafting density, above which highly ordered hexagonal structures are achieved, as illustrated in Figure \ref{fig:DLS_TGA_main} (b).

\begin{table*}[ht]
\caption{Summary of the lattice constants ($a$), nearest-neighbor distances ($d_{\rm NN}$), and FWHM for different PEG-AgNP assemblies induced under various electrolytes as indicated.}
\label{table:PEG_conditions}
\begin{threeparttable}
\normalsize
\centering
    \begin{tabular}{cllcccc}
        \midrule \midrule
        MW of grafted & Solvent & Lattice  & $a\tnote{a}$ & $d_{\rm NN}\tnote{b}$ & FWHM$\tnote{c}$ & $\Gamma _{\rm e}\tnote{d}$\\ 
        PEG (kDa) & Condition &Type & (nm) & (nm) & (\AA$^{-1}$) & (e/\AA$^{2}$) \\
        \midrule 
        \multirow{8}{*}{2} & Water & n.a. & $--$ & $--$ & $--$ &\textless\ 0.5\\ 
        & \ch{K2CO3} 1 mM & MRO (1D like) & n.a.&$\sim$ 34 & $\sim$ 0.012 &$\sim$ 7\\ 
        & \ch{K2CO3} 10 mM & MRO (1D like)  & n.a.&$\sim$ 26 & $\sim$ 0.006& $\sim$ 14\\ 
        & \ch{K2CO3} 100 mM & Hexagonal & $\sim$ 24&$\sim$ 24 & $\sim$ 0.004 &$\sim$ 29\\ 
        & PAA 0.2 mM & n.a. & $--$ & $--$ & $--$ &\textless\ 0.5\\ 
        & PAA 2 mM & n.a. & $--$ & $--$ & $--$ &\textless\ 0.5\\ 
        & \multirow{2}{*}{\makecell[l]{PAA 2 mM $+$\\10 mM HCl}} & \multirow{2}{*}{MRO (Hexagonal)} & \multirow{2}{*}{$\sim$ 20} & \multirow{2}{*}{$\sim$ 20} & \multirow{2}{*}{$\sim $0.004} &\multirow{2}{*}{$\sim$ 2}\\ 
        & & & & & &\\
        
        \midrule 
        \multirow{8}{*}{5} & Water & n.a. & $--$ & $--$ & $--$ &\textless\ 0.5\\ 
        & \ch{K2CO3} 1 mM & Hexagonal & $\sim$ 40&$\sim$ 40 & $\sim$ 0.003 &$\sim$ 9\\ 
        & \ch{K2CO3} 10 mM & Hexagonal & $\sim$ 36&$\sim$ 36 & $\sim$ 0.003 &$\sim$13 \\ 
        & \ch{K2CO3} 100 mM & Hexagonal & $\sim$ 36&$\sim$ 36 & $\sim$ 0.003 &$\sim$ 15 \\ 
        & PAA 0.2 mM & n.a. & $--$ & $--$ & $--$ &$\sim$ 3\\ 
        & PAA 2 mM & n.a. & $--$ & $--$ & $--$ &$\sim$ 4\\ 
        & \multirow{2}{*}{\makecell[l]{PAA 2 mM $+$\\10 mM HCl}} & \multirow{2}{*}{Hexagonal} & \multirow{2}{*}{$\sim$ 29} & \multirow{2}{*}{$\sim$ 29} & \multirow{2}{*}{$\sim$ 0.003} &\multirow{2}{*}{$\sim$ 8}\\ 
        & & & & & &\\

        \midrule 
        \multirow{8}{*}{20} & Water & SRO & n.a.&$\sim$ 72 & $\sim$ 0.006 &$\sim$ 1\\
        & \ch{K2CO3} 1 mM & SRO & n.a.&$\sim$ 63 & $\sim$ 0.006 &$\sim$ 1\\
        & \ch{K2CO3} 10 mM & SRO & n.a.&$\sim$ 69 & $\sim$ 0.006 &$\sim$ 1 \\
        & \ch{K2CO3} 100 mM & SRO & n.a.&$\sim$ 72 & $\sim$ 0.006 &$\sim$ 2\\
        & PAA 0.2 mM & SRO & n.a.&$\sim$ 78 & $\sim$ 0.007 &$\sim$ 2\\ 
        & PAA 2 mM & SRO & n.a.&$\sim$ 69 & $\sim$ 0.007 &$\sim$ 3\\ 
        & \multirow{2}{*}{\makecell[l]{PAA 2 mM $+$\\10 mM HCl}} & \multirow{2}{*}{MRO (Hexagonal)} & \multirow{2}{*}{$\sim$ 37} & \multirow{2}{*}{$\sim$ 37} & \multirow{2}{*}{$\sim$ 0.005} &\multirow{2}{*}{$\sim$ 5}\\ 
        & & & & & &\\
        
        \midrule 
        \multirow{8}{*}{40} & Water & SRO (Hexagonal) & $\sim$ 83 &$\sim$ 83 & $\sim$ 0.002 &$\sim$ 1\\
        & \ch{K2CO3} 1 mM & SRO (Hexagonal) & $\sim$ 76 &$\sim$ 76 & $\sim$ 0.002 &$\sim$ 1\\
        & \ch{K2CO3} 10 mM & SRO (Hexagonal) & $\sim$ 70 &$\sim$ 70 & $\sim$ 0.002 &$\sim$ 2\\
        & \ch{K2CO3} 100 mM & SRO (Hexagonal) & $\sim$ 70 &$\sim$ 70 & $\sim$ 0.002 &$\sim$ 2\\
        & PAA 0.2 mM & SRO (Hexagonal) & $\sim$ 82&$\sim$ 82 & $\sim$ 0.001 &$\sim$ 3 \\ 
        & PAA 2 mM & SRO (Hexagonal) & $\sim$ 72&$\sim$ 72 & $\sim$ 0.002 &$\sim$ 5\\ 
        & \multirow{2}{*}{\makecell[l]{PAA 2 mM $+$\\10 mM HCl}} & \multirow{2}{*}{\makecell[l]{SRO (Hexagonal)$^*$\\Hexagonal$^*$}} & \multirow{2}{*}{\makecell[l]{$\sim$ 59\\$\sim$ 42}} & \multirow{2}{*}{\makecell[l]{$\sim$ 59\\$\sim$ 42}} & \multirow{2}{*}{\makecell[l]{$\sim$ 0.005\\$\sim$ 0.002}} &\multirow{2}{*}{$\sim$ 9}\\ 
        & & & & & &\\
        \midrule \midrule
    \end{tabular}
\begin{tablenotes}
    \item[a] {\scriptsize For 2D hexagonal structure, $a = 4 \pi /(\sqrt{3} Q_1)$.}
    \item[b] {\scriptsize For 2D hexagonal structure, $d_{\rm NN}= 4 \pi /(\sqrt{3} Q_1)$. For MRO and SRO, $d_{\rm NN}= 2 \pi /Q_1$.}
    \item[c] {\scriptsize FWHM - Full Width at Half Maximum is estimated from the Lorentzian-fitting function of the first peak.}
    \item[d] {\scriptsize The excess of surface electron density in ED profiles, $\Gamma_{\rm e}=\int[\rho(z)-\rho_{\rm sub}(z)]{\rm d}z$.}
    \item[*] {\scriptsize The structure in the presence of 2 mM PAA and 10 mM HCl sub-phase condition is separated. Both the phases are hexagonal and structural parameters are listed in the particular row.}
\end{tablenotes}
\end{threeparttable}
\end{table*}

\begin{table*}[htbp]
\caption{Summary of the lattice constants ($a$), nearest-neighbor distances ($d_{\rm NN}$), and FWHM for various individual Au/AgNP and 1:1 binary mixtures of PEG-AgNP and AuNP assemblies formed in 100 mM \ch{K2CO3}.}
\label{bianry_table}
\begin{threeparttable}
\centering
\begin{tabular}{llcccccc}
\toprule
\multirow{3}{*}{\textbf{NP1}} & \multirow{3}{*}{\textbf{NP2}} & \multirow{3}{*}{\textbf{Lattice type}} & \multicolumn{4}{c}{\textbf{Lattice Parameters}} \\
& & & $a\tnote{a}$ & $d_{\rm NN}\tnote{b}$ & FWHM$\tnote{c}$ & $\Gamma_{\rm e}\tnote{d}$ \\
& & & (nm) & (nm) & (\AA$^{-1}$) & (e/\AA$^{2}$) \\
\midrule
PEG2k-AgNP10         &         $--$            & Hexagonal            & $\sim$25        & $\sim$25          & $\sim$0.002       & $\sim$35          \\
       $--$              & PEG5k-AuNP5         & Hexagonal            & $\sim$29        & $\sim$29          & $\sim$0.001       & $\sim$9           \\
PEG2k-AgNP10         & PEG5k-AuNP5         & Hexagonal$^*$        & $\sim$29        & $\sim$29          & $\sim$0.002       & $\sim$12          \\
  $--$              & PEG5k-AuNP10        & Hexagonal            & $\sim$27        & $\sim$27          & $\sim$0.001       & $\sim$22          \\
PEG2k-AgNP10         & PEG5k-AuNP10        & Hexagonal            & $\sim$30        & $\sim$30          & $\sim$0.002       & $\sim$24          \\
PEG5k-AgNP10         &         $--$          & Hexagonal            & $\sim$36        & $\sim$36          & $\sim$0.002       & $\sim$18          \\
PEG5k-AgNP10         & PEG5k-AuNP5         & Hexagonal (SRO)      & $\sim$36        & $\sim$36          & $\sim$0.003       & $\sim$8           \\
\bottomrule
\end{tabular}
\begin{tablenotes}
    \item[a] {\scriptsize For 2D hexagonal structure, $a = 4 \pi /(\sqrt{3} Q_1)$.}
    \item[b] {\scriptsize For 2D hexagonal structure, $d_{\rm NN}= 4 \pi /(\sqrt{3} Q_1)$.}
    \item[c] {\scriptsize FWHM - Full Width at Half Maximum is estimated from the Lorentzian-fitting function of the first peak.}
    \item[d] {\scriptsize The excess of surface electron density in ED profiles, $\Gamma_{\rm e}=\int[\rho(z)-\rho_{\rm sub}(z)]{\rm d}z$.}
    \item[*] {\scriptsize The hexagonal ordering formed here is heterogeneous.}
\end{tablenotes}
\end{threeparttable}
\end{table*}

\section{Conclusions}

In this study, we successfully grafted PEG onto AgNP surfaces and investigated the effects of PEG chain length, electrolyte environment, and grafting density on the structural ordering of PEG-grafted AgNPs. The grafting density of thiolated PEG is lower for AgNPs compared to AuNPs. Using synchrotron-based GISAXS and XRR measurements, we examined the in-plane ordering and surface distribution of nanoparticles at the vapor-liquid interface under different suspension conditions. Previous studies on AgNPs with a 20 nm core in \ch{K2CO3} have shown the formation of a 2D hexagonal structure at the liquid/vapor interface, while the addition of PAA induces a transition to a lamellar-like structure. It has been suggested that the lower grafting density of AgNPs, compared to AuNPs, causes AgNPs to form chains that deviate from ideal hexagonal packing, resulting in less ordered structures. Our findings using 10 nm core and including shorter PEG chain length (2 kDa) reveal that shorter PEG chains lead to well-ordered hexagonal structures with higher surface densities for both AgNPs and AuNPs, while longer chains induce phase separation and result in less dense quasi-bilayer structures. This behavior is driven by steric hindrance and interparticle electrostatic interactions. We identify a critical grafting density, regardless of particle type, that marks the threshold between highly ordered NPs and poorly ordered ones.
When exploring binary systems of PEG-AuNPs and PEG-AgNPs, the results show hybrid structures with disrupted ordering. In most cases, AuNPs dominate the surface arrangement, while AgNPs fuse into the AuNP- matrix, leading to expanded lattice constants but lacking the ordering seen in individual systems.

Although this study offers important insights into nanoparticle assembly, further work is needed to achieve more ordered binary superstructures. Future investigations should focus on varying AgNP concentrations and exploring the use of charged PEG chains to induce ionic interactions and superlattice formation. These strategies could provide better control over the structural ordering of binary nanoparticle systems, enhancing their potential for various applications. 

\section{Acknowledgements}
The research was supported by the U.S. Department of Energy (U.S. DOE), Office of Basic Energy Sciences, Division of Materials Sciences and Engineering. J Ethan Batey thanks the Science Undergraduate Laboratory Internships Program (SULI) for supporting the conduction of research at Ames National Laboratory. Iowa State University operates Ames Laboratory for the U.S. DOE under Contract DE-AC02-07CH11358. Part of this research used the Open Platform Liquid Scattering (OPLS) end station of the Soft Matter Interfaces Beamline (SMI, Beamline 12-ID)  at the National Synchrotron Light Source II (Brookhaven National Laboratory), a US Department of Energy (DOE) Office of Science User Facility operated for the DOE Office of Science by Brookhaven National Laboratory under Contract No. DE-SC0012704.  We thank Dr. Benjamin M. Ocko for his support in X-ray scattering experiments at OPLS. Part of this research used NSF’s ChemMatCARS Sector 15. NSF’s ChemMatCARS Sector 15 is supported by the Divisions of Chemistry (CHE) and Materials Research (DMR), National Science Foundation, under grant number NSF/CHE-1834750. The use of the Advanced Photon Source, an Office of Science User Facility operated for the U.S. Department of Energy (DOE) Office of Science by Argonne National Laboratory, was supported by the U.S. DOE under Contract No. DE-AC02-06CH11357. 

\section{Author contributions}
WW, DV, and SM conceived and supervised the project. BN, HJK, DV, and WW designed and conducted the experiments. BN, JEB, DV, and WW analyzed the data. BN, WW, and DV wrote the manuscript. WB and HZ supported X-ray scattering experiments, data acquisition, and data processing. SM, DV, and WW secured funding for the project.  All co-authors read and reviewed the manuscript. 

\section{Supporting Information}
The Supporting Information is available free of charge on the
Publishers website at DOI: xxxxx/yyyyy

Supporting Information: DLS and Thermo-gravimetric analysis; Additional PEG-AgNPs data; Additional binary mixture data and analysis

\section{Data Availability}
The findings of this study are supported by data found within the article and the provided Supplementary Information. Further relevant information and source data can be obtained from the corresponding author upon making a reasonable request.

\section*{Additional information}
The corresponding author declares that they have no competing financial interests on behalf of all authors of the paper. 

\clearpage
\normalem
\bibliography{Reference.bib}

\onecolumn
 
\clearpage
\onecolumn

\setcounter{page}{1}
\setcounter{figure}{0}
\setcounter{equation}{0}
\setcounter{table}{0}

\renewcommand{\thefigure}{S\arabic{figure}}
\renewcommand{\theequation}{S\arabic{equation}}
\renewcommand{\thetable}{S\arabic{table}}
\renewcommand{\thepage}{S\arabic{page}} 

\section{{\Large Supporting information}}
\begin{center} 
	{\bf \Large Effect of Grafting Density on the Two-dimensional Assembly of Nanoparticles}\\
	\bigskip
	\normalsize
    Binay P. Nayak,$^\dagger$
    James Ethan Batey, $^{\ddagger,\perp}$ 
	Hyeong Jin Kim,$^\dagger$ 
	Wenjie Wang,$^\ddagger$
    Wei Bu $^\mathparagraph$ 
	Honghu Zhang,$^{\S,\#}$ 
	Surya K. Mallapragada,$^*$$^,$$^\dagger$ and David Vaknin$^*$$^,$$^{\mid\mid}$\\
	\bigskip
	{$\dagger$\it Ames National Laboratory, and Department of Chemical and Biological Engineering, Iowa State University, Ames, Iowa 50011, United States}\\
	{$\ddagger$\it Division of Materials Sciences and Engineering, Ames National Laboratory, U.S. DOE, Ames, Iowa 50011, United States}\\
    {$\mathparagraph$\it NSF’s ChemMatCARS, Pritzker School of Molecular Engineering, University of Chicago, Chicago, Illinois 60637, United States}\\
	{$\S$\it Center for Functional Nanomaterials, Brookhaven National Laboratory, Upton, New York 11973, United States}\\
	{$\#$\it National Synchrotron Light Source II, Brookhaven National Laboratory, Upton, New York 11973, United States}\\
	{$\mid \mid$\it Ames National Laboratory, and Department of Physics and Astronomy, Iowa State University, Ames, Iowa 50011, United States}\\
    {$\perp$\it Department of Chemistry and Biochemistry, University of Arkansas, Fayetteville, Arkansas 72701, United States}\\
	\bigskip
	{E-mail: suryakm@iastate.edu; vaknin@ameslab.gov}\\
\end{center}

\subsection{DLS and Thermo-gravimetric analysis}
\normalsize
\begin{figure}[!ht]
  \centering
  \begin{minipage}{.45\textwidth}
        \centering 
 	\includegraphics[width=\linewidth]{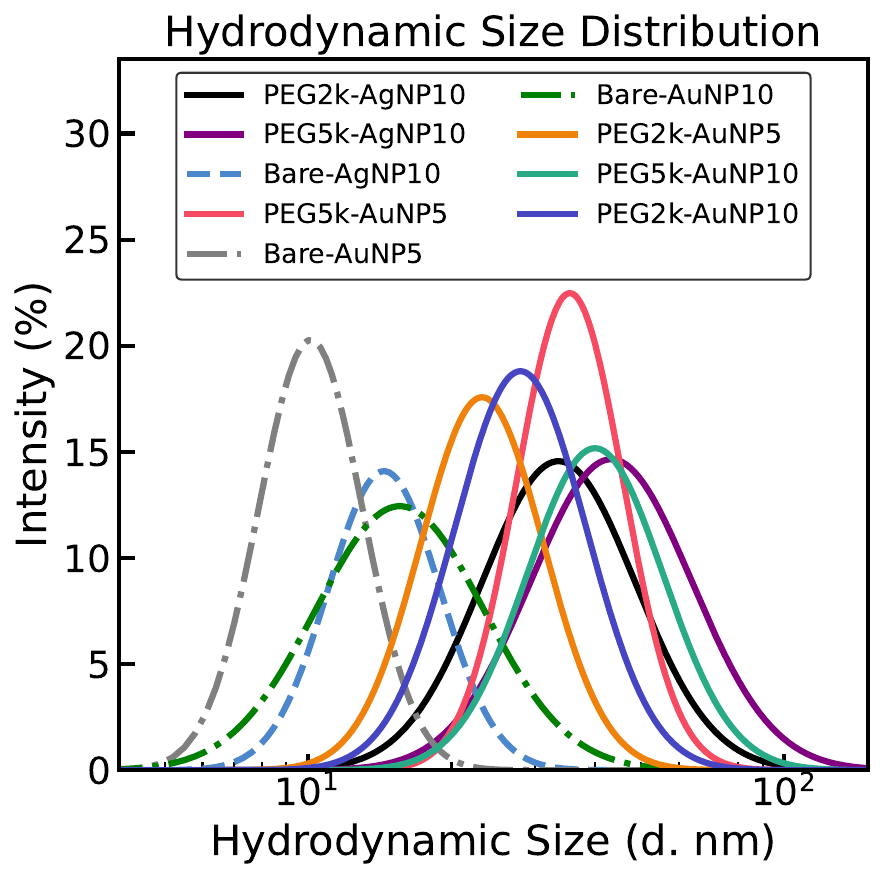}
 	 \caption{\scriptsize DLS intensity percentage versus hydrodynamic size distribution for aqueous suspensions of nanoparticles. Bare surface AgNPs are represented by a dashed line (\raisebox{0.5ex}{\rule{0.5em}{0.4pt}}\hspace{0.2em}\raisebox{0.5ex}{\rule{0.5em}{0.4pt}}), AuNPs by a dash-dot line (\raisebox{0.5ex}{\rule{0.5em}{0.4pt}}\hspace{0.2em}$\cdot$\hspace{0.2em}\raisebox{0.5ex}{\rule{0.5em}{0.4pt}}), and PEG-grafted AuNPs and AgNPs by a solid line (\rule[0.5ex]{1em}{0.4pt}).}
 	\label{fig:DLS}
  \end{minipage}%
    \hspace{0.05\textwidth} 
    \begin{minipage}{.45\textwidth}
 	\centering 
 	\includegraphics[width=\linewidth]{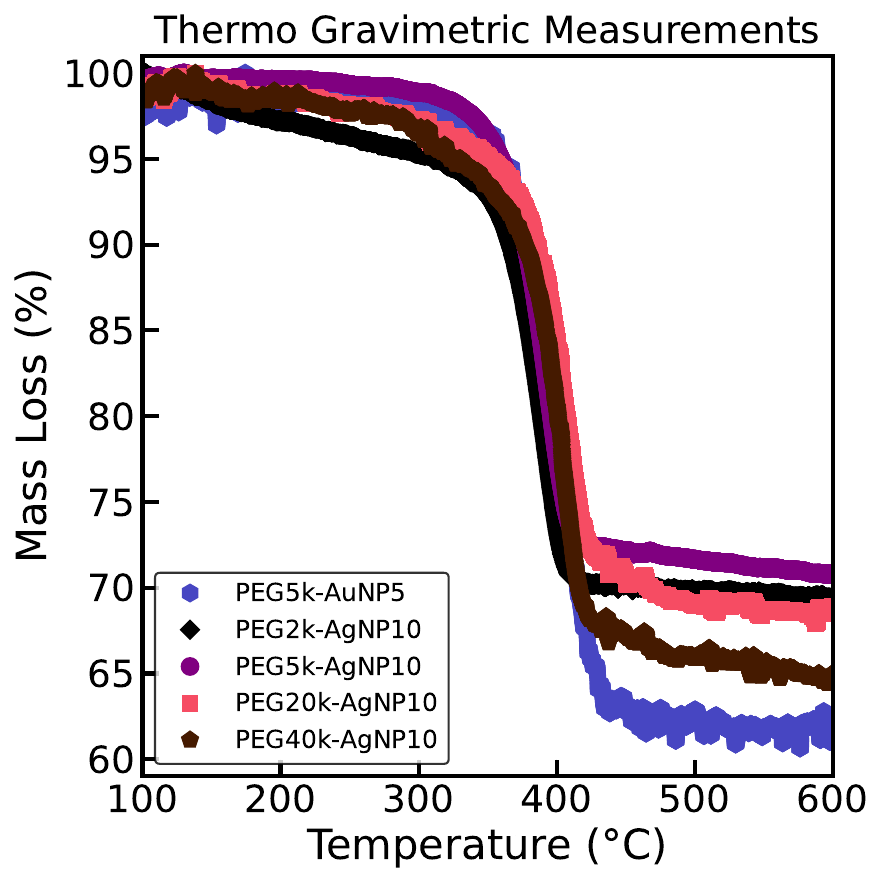}
    \caption{\scriptsize TGA graphs of PEG-grafted AgNPs and AuNPs, showing the thermal degradation of functionalized PEG on nanoparticle surfaces between 350 \textdegree{C} and 450 \textdegree{C}. The calculated $\sigma$ values and initial weights are summarized in Table \ref{tbl:dls-tga}.}
 	\label{fig:TGA}
  \end{minipage}
 \end{figure}
 
Figure \ref{fig:DLS} presents the hydrodynamic size (\(D_{\rm H}\)) distribution for aqueous suspensions of nanoparticles (NPs) used in this study, both before and after PEG grafting. As expected, the hydrodynamic size increases with the PEG chain length (or molecular weight), reflecting successful PEG grafting. The summarized results can be found in Table \ref{tbl:dls-tga}. The \(D_{\rm H}\) for bare NPs is slightly larger than their nominal sizes ($\sim$ 10 nm or 5 nm). This increase is likely due to the stabilization by citrate ions, which are adsorbed onto the AuNP surfaces, thereby increasing the hydrodynamic size compared to the nominal size. It is also noteworthy that PEG2k-AgNPs and PEG5k-AuNPs have comparable \(D_{\rm H}\), which makes them suitable candidates for use as components in a binary system.

Thermogravimetric analysis (TGA) values of PEG-grafted AgNPs and AuNPs, showing the thermal degradation of functionalized PEG on nanoparticle surfaces between 350 \textdegree{C} and 450 \textdegree{C} under an argon atmosphere. The estimated grafting densities for PEG2k-AgNP, PEG5k-AgNP, PEG20k-AgNP, and PEG40k-AgNP are approximately 1.85, 0.87, 0.22, and 0.14 PEG chains per nm\(^2\) per nanoparticle (NP), respectively. The summarized results can be found in Table \ref{tbl:dls-tga}. Given that PEG2k, PEG5k, PEG20k, and PEG40k consist of roughly 45, 114, 454, and 908 PEG monomers, their grafting densities correspond to 83, 99, 100, and 127 PEG monomers per nm\(^2\) per NP, respectively. For PEG5k-AuNP, the grafting density is approximately 2.50 chains per nm\(^2\) per NP or 285 monomers per nm\(^2\) per NP.

\subsection{Additional PEG-AgNPs data}

\begin{figure}
    \centering
    \includegraphics[width=0.95 \linewidth]{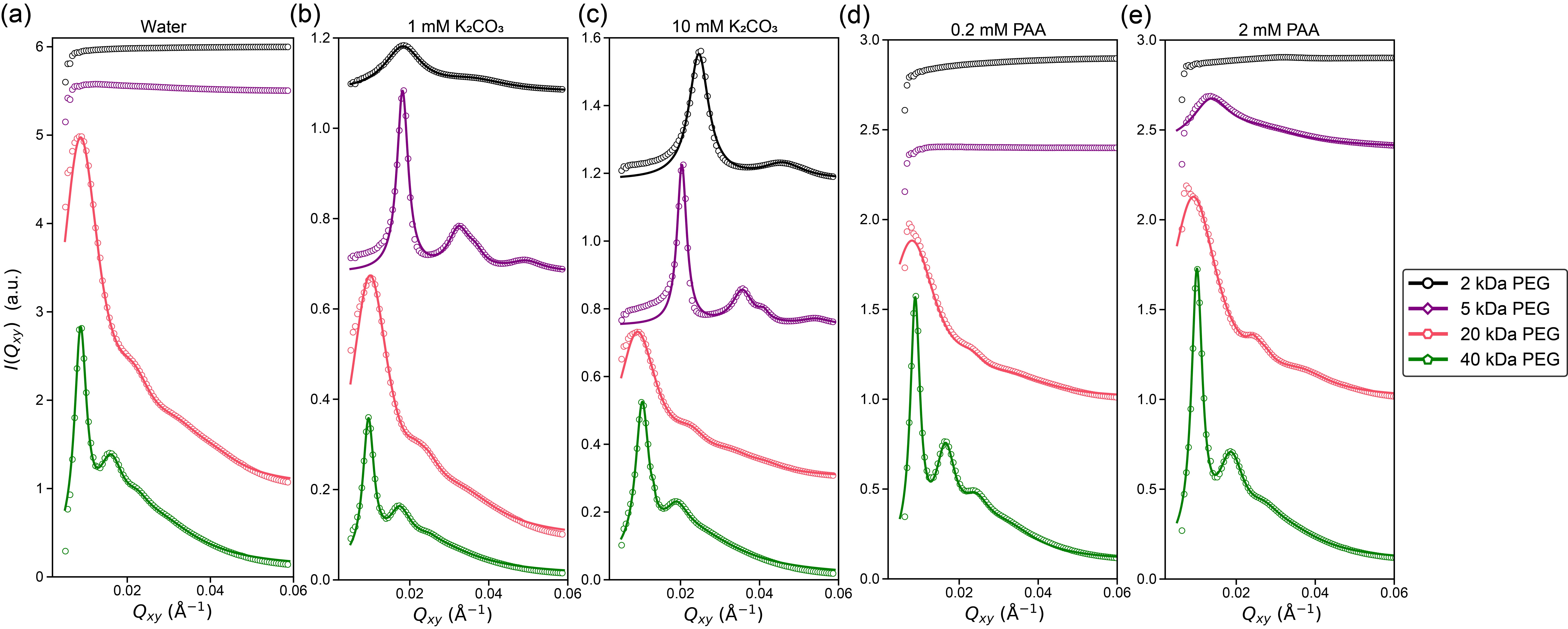}
    \caption{GISAXS linecut profiles displaying intensity ($I(Q_{xy})$) versus in-plane momentum transfer ($Q_{xy}$) for PEG-AgNPs with various MW in suspensions. (a) in water (b) in 1 mM \ch{K2CO3} (c) in 10 mM \ch{K2CO3} (d) in 0.2 mM PAA (e)in 2 mM PAA. The solid lines represent the best-fit Lorentzian profiles to the experimental data. The plots are vertically shifted for clarity, and in some cases, the intensity of the plots is multiplied by a constant to enhance visibility.}
    \label{fig:gsx_all}
\end{figure}

\begin{figure}[!ht]
  \centering
  \begin{minipage}{.45\textwidth}
        \centering 
        \includegraphics[width=\linewidth]{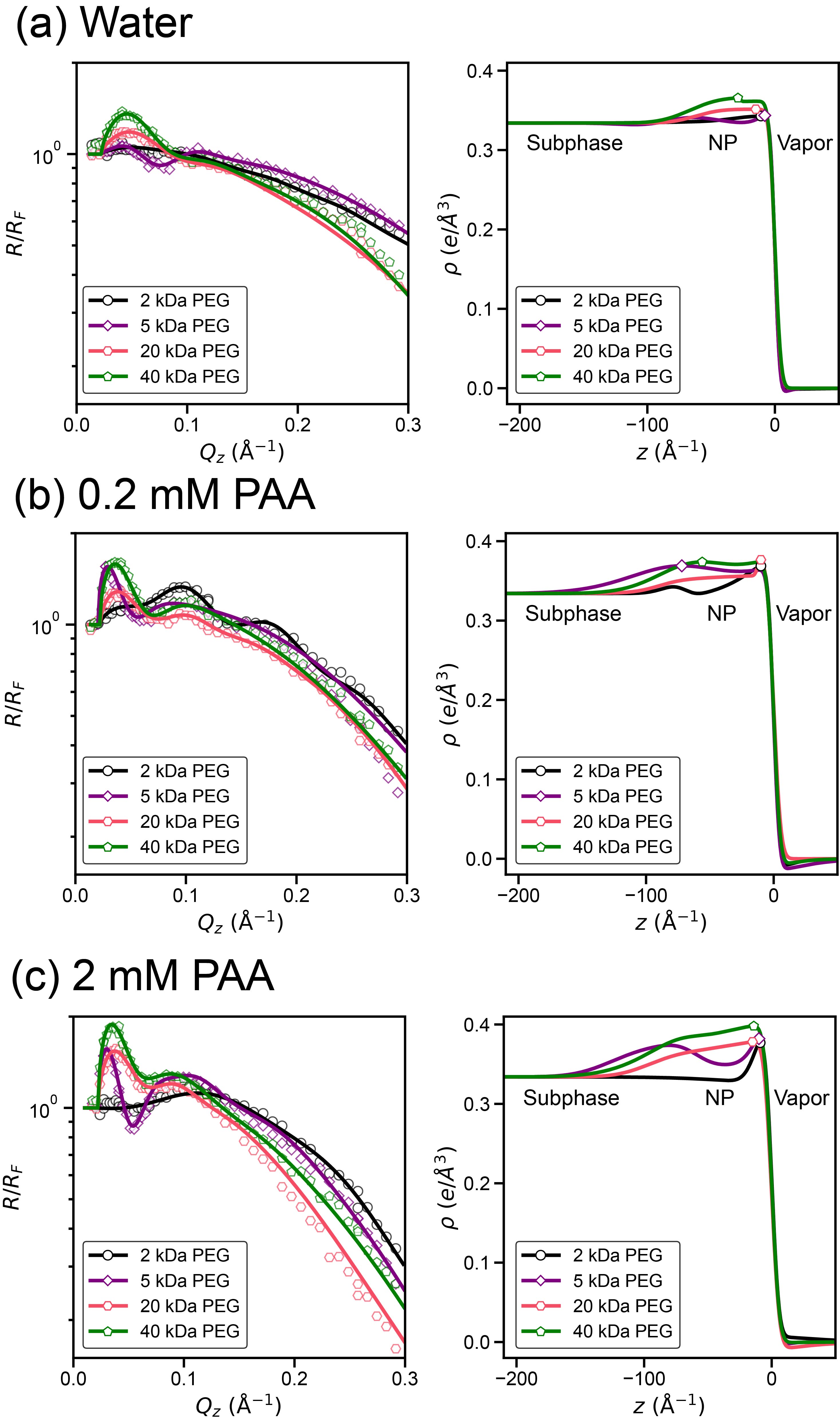}
        \caption{Normalized X-ray reflectivity profiles \((R/R_{\text{F}})\) for PEG-AgNPs in different environments: (a) water, (b) 0.2 mM PAA, and (c) 2 mM PAA, with varying PEG MW as indicated. The corresponding ED profiles are shown to the right of each plot.}
        \label{fig:ref_all}
  \end{minipage}%
  \hspace{0.05\textwidth} 
  \begin{minipage}{.45\textwidth}
        \centering 
        \includegraphics[width=\linewidth]{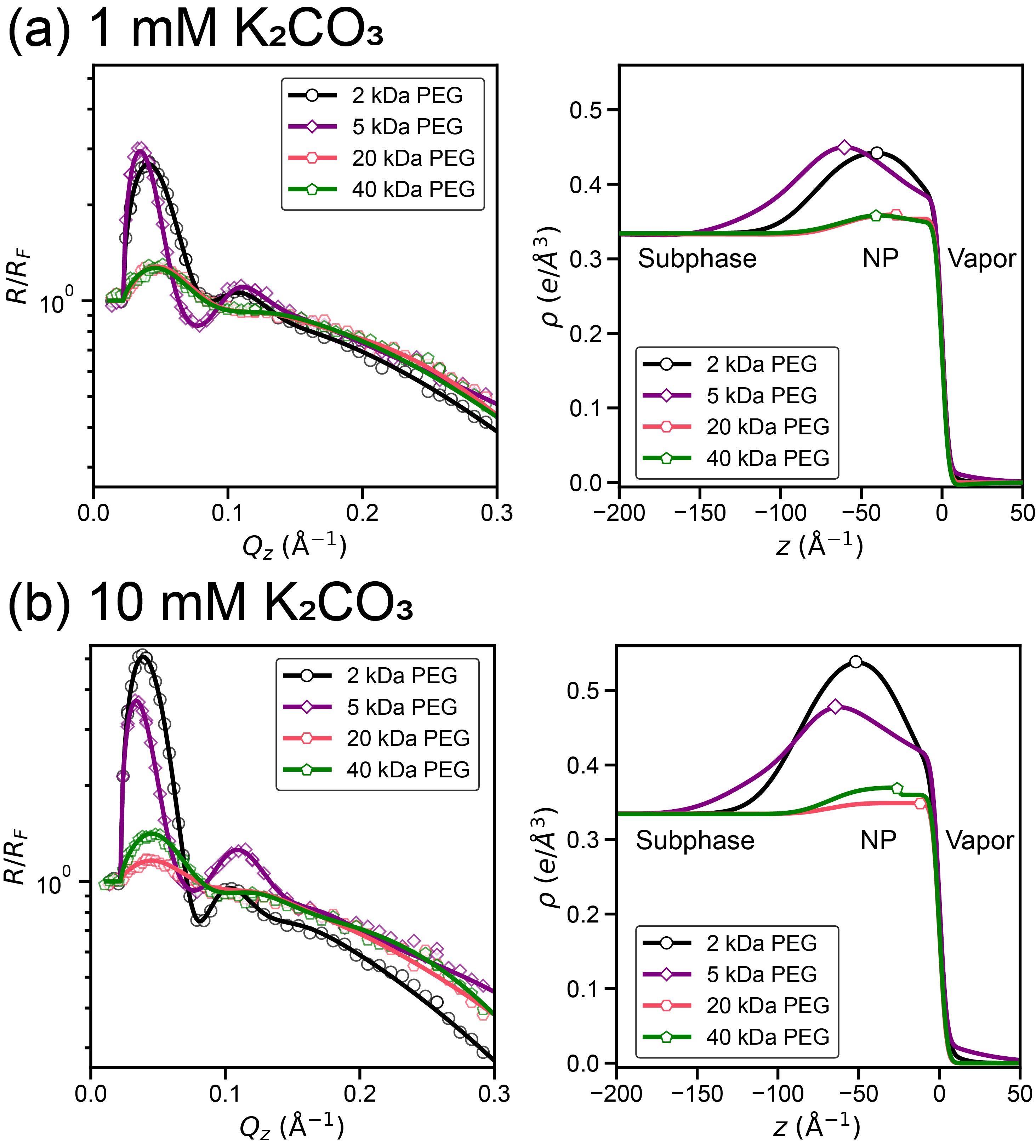}
        \caption{Normalized X-ray reflectivity profiles \((R/R_{\text{F}})\) for PEG-AgNPs in different environments: (a) 1 mM \ch{K2CO3}, (b) 10 mM \ch{K2CO3}, with varying PEG MW as indicated. The corresponding ED profiles are shown to the right of each plot.}
        \label{fig:ref_k}
  \end{minipage}
\end{figure}

Figure \ref{fig:ref_all} and Figure \ref{fig:ref_k} present normalized X-ray reflectivity profiles \((R/R_{\text{F}})\) for PEG-AgNPs under different environmental conditions, as indicated. The best-fit solid lines, derived from the ED profiles, are displayed to the right of each plot. A systematic increase in electrolyte concentration and corresponding measurements were conducted. In the presence of \ch{K2CO3}, the XRR data show the formation of single-layer films, while in the presence of PAA, with or without HCl, incomplete or quasi-bilayer films are observed.

Figure \ref{fig:gsx_all} shows GISAXS results, displaying the intensity ($I(Q_{xy})$) versus in-plane momentum transfer ($Q_{xy}$) for PEG-AgNPs with various MW in suspensions, as indicated. The solid lines represent the best-fit Lorentzian profiles for the experimental data. Without any electrolyte, PEG40k-AgNPs and PEG20k-AgNPs exhibit short-range ordering, which remains unaltered with increasing \ch{K2CO3} concentrations up to 10 mM, as well as with the addition of PAA up to 2 mM. The structural parameters obtained from the GISAXS data and the surface-excess electron density from the XRR data are summarized in Table \ref{table:PEG_conditions}.

\subsection{Additional binary mixture data and analysis}

\begin{figure*}[!hbt]
    \centering
    \includegraphics[width=1\linewidth]{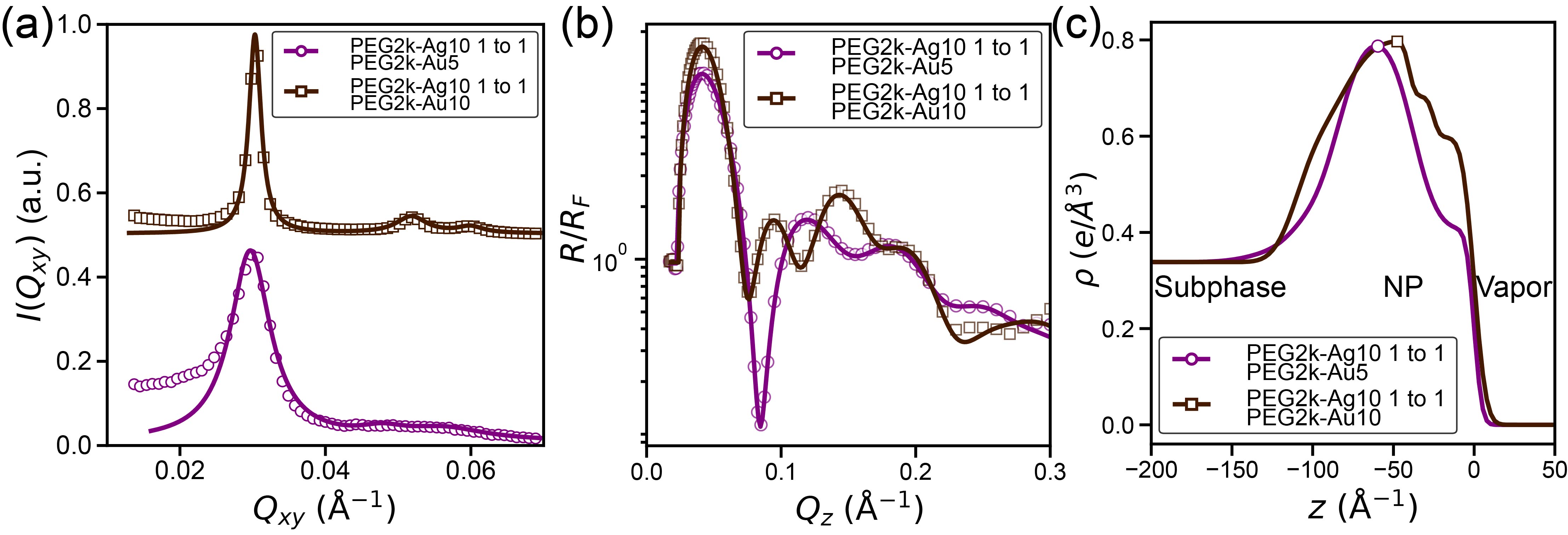}
    \caption{\scriptsize (a) GISAXS line-cut profiles showing intensity \((I(Q_{xy}))\) versus in-plane momentum transfer \((Q_{xy})\) for a binary mixture of PEG2k-Ag10 and PEG2k-Au5 ($\bigcirc$), and PEG2k-Ag10 and PEG2k-Au10 ($\square$). The solid lines represent the best-fit Lorentzian profiles to the experimental data. (b) Normalized X-ray reflectivity profiles \((R/R_{\text{F}})\) for the same films depicted in (a), with the best-fit solid line obtained from the ED profiles shown in (c).}
    \label{21025_gsx}
\end{figure*}

Figure \ref{21025_gsx} presents GISAXS line-cut profiles showing intensity \((I(Q_{xy}))\) versus in-plane momentum transfer \((Q_{xy})\) for binary mixtures of PEG2k-Ag10 and PEG2k-Au5 ($\bigcirc$), and PEG2k-Ag10 and PEG2k-Au10 ($\square$). The solid lines represent the best-fit Lorentzian profiles from the experimental data. Panel (b) shows the normalized X-ray reflectivity profiles \((R/R_{\text{F}})\) for the same films depicted in (a), with the best-fit solid lines obtained from the ED profiles shown in (c). The GISAXS data indicate the formation of a well-ordered hexagonal structure for the PEG2k-Ag10 and PEG2k-Au5 mixture, predominantly formed by the AuNPs. In contrast, the mixture of PEG2k-Ag10 and PEG2k-Au10 shows the formation of short-range disordered structures. The XRR data, along with the ED profiles, suggest the formation of complete single-layer films at the interface, which may be hybrid in nature due to the higher ED values observed in the plot.

\begin{figure*}[!hbt]
    \centering
    \includegraphics[width=1\linewidth]{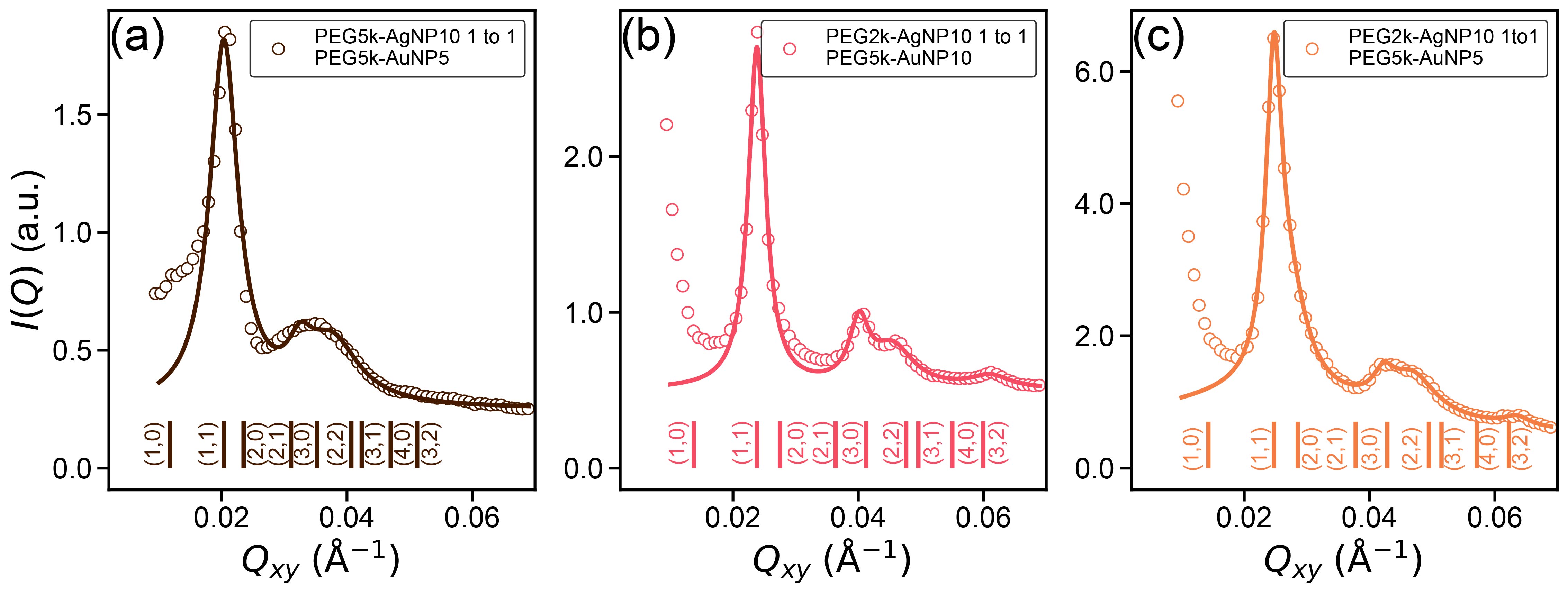}
    \caption{\scriptsize (a) GISAXS line-cut profiles showing intensity \((I(Q_{xy}))\) versus in-plane momentum transfer \((Q_{xy})\) for 1:1 binary mixtures of PEG5k-AgNP10 \& PEG5k-AuNP5, PEG2k-AgNP10 \& PEG5k-AuNP10, and PEG2k-AgNP10 \& PEG5k-AuNP5. The solid lines represent the best-fit Lorentzian profiles to the experimental data. The vertical bars indicate the calculated peak positions for a \(\sqrt{3} \times \sqrt{3}\) hexagonal lattice (AB$_2$ type), with the prominent peak corresponding to the (1,1) plane. The peak positions are marked with the respective Miller indices.}
    \label{binary_peaks}
\end{figure*}

Figure \ref{binary_peaks} shows the calculated peak positions for a \(\sqrt{3} \times \sqrt{3}\) hexagonal lattice. However, the GISAXS patterns indicate that some of the calculated peaks are missing, which could be attributed to the formation of an incomplete or hybrid lattice in the binary mixture of Au and AgNPs.

\end{document}